\documentclass[openac]{rsproca_new}

\usepackage{amsmath}
\usepackage{bm}
\usepackage{dsfont}
\usepackage{graphicx}
\usepackage{enumitem}
\usepackage{color}
\usepackage{mathrsfs}

\usepackage{newtxtext,newtxmath}

\definecolor{OliveGreen}{rgb}{0.1, 0.4, 0.1}

\newcommand{\ch}{\operatorname{ch}}
\newcommand{\sh}{\operatorname{sh}}
\renewcommand{\th}{\operatorname{th}}
\newcommand{\cth}{\operatorname{cth}}

\newcommand{\csh}{\operatorname{csh}} 
\newcommand{\sch}{\operatorname{sch}} 
\newcommand{\K}{\mathrm{K}}

\usepackage{bigints}

\usepackage[left = 2.2cm, right = 2.2cm, 
top = 2.6cm, bottom = 2.6cm]{geometry}

\allowdisplaybreaks

\definecolor{awesome}{rgb}{0.9, 0.3, 0.2}

\titlehead{Research}

\begin{document}

\title{Phoretic flow in a three-dimensional wedge geometry}

\author{
Abdallah Daddi-Moussa-Ider$^{1}$,
Semyon Yakubovich$^{2}$,
Maciej Lisicki$^{3}$
}

\address{$^{1}$School of Mathematics and Statistics, The Open University, Walton Hall, Milton Keynes MK7 6AA, United Kingdom\\
$^{2}$Department of Mathematics, Faculty of Sciences, University of Porto, Rua do Campo Alegre 1021/1055, 4169-007 Porto, Portugal\\
$^{3}$Faculty of Physics, University of Warsaw, Pasteura~5, Warsaw 02-093, Poland}

\subject{applied mathematics, biophysics, fluid
mechanics}


\corres{A. Daddi-Moussa-Ider\\
\email{admi2@open.ac.uk}}

\begin{abstract}
Understanding how chemically induced surface transport generates fluid motion in confined geometries is essential for the rational design of microscale pumping devices and active microfluidic systems.
Here we develop a theoretical framework for chemically driven phoretic flows in a three-dimensional wedge geometry in the diffusion-dominated regime. We formulate both the diffusion and hydrodynamic problems using a Fourier–Kontorovich–Lebedev spectral representation, exploiting the translational invariance and radial structure of the wedge. Green's functions for the concentration field are derived for reflecting and mixed reflecting–absorbing boundaries, reducing to finite image-like sums or closed-form expressions for commensurate wedge angles. The resulting slip velocity is then used to construct the three-dimensional Stokes flow through the Papkovich–Neuber representation, yielding explicit spectral solutions for the velocity field. These results establish a Green's-function framework for phoretic pumping in wedge-shaped confinement and provide analytical benchmarks for numerical simulations of chemically driven transport in confined microfluidic systems.

\end{abstract}

\maketitle

\section{Introduction}

Microscale flows are dominated by viscous effects and are therefore widely known to be difficult to stir or pump locally because the absence of inertia prevents the spontaneous generation of large-scale mixing mechanisms. Boundary forcing and interfacial physicochemical effects provide an attractive alternative route for generating localised streaming and pumping at small scales~\cite{squires2005,anderson1989,Shim2022}. Among these mechanisms, diffusioosmosis and diffusiophoresis offer a means of converting chemical gradients into fluid motion through solute--surface interactions within a thin interfacial layer adjacent to a boundary. When the characteristic thickness of this layer is small compared with the relevant flow length scales, its net effect can be represented by an effective tangential slip velocity at the surface~\cite{Derjaguin1993,anderson1984,anderson1989}, analogous to the surface-driven flows generated by ciliated microswimmers~\cite{blake71cilia, brennen77}.

For chemically active surfaces, solute gradients arise from local production or consumption mechanisms, such as catalytic reactions, which are commonly modelled through a prescribed normal solute flux at the boundary. The resulting tangential concentration gradients generate a diffusioosmotic slip velocity that acts as an effective boundary condition for the bulk Stokes flow. This slip velocity is proportional to the tangential gradient of the solute concentration along the surface. In the limit of small solute Péclet number, where advection by the induced flow is negligible compared with molecular diffusion, the chemical and hydrodynamic problems decouple. The concentration field can then be determined independently from a diffusion problem with boundary conditions set by the surface chemistry, and the resulting slip distribution provides the forcing for the fluid motion. Through the Lorentz reciprocal theorem~\cite{masoud2019reciprocal}, the prescribed slip velocity is directly related to the translational and rotational motion of microswimmers~\cite{stone96}. This framework underpins the modern continuum description of chemically driven colloids, including the autophoretic motion of synthetic swimmers such as active colloids~\cite{bechinger16,varma2019modeling,daddi2022diffusiophoretic}, as well as phoretic pumping systems~\cite{julicher2009,sabass2012,michelin2014}. The pioneering formulation of Golestanian and collaborators~\cite{golestanian2007} established the theoretical foundations and design principles for artificial phoretic swimmers.

The dual problem to microswimming is microscale boundary-driven pumping, where phoretic mechanisms are exploited to generate directed flows in confined domains~\cite{golestanian2007,michelin2015autophoretic,Lisicki2018,michelin2015geometric,michelin2019universal,Yu2020}. Such flows often arise in strongly confined geometries, where nearby boundaries enhance phoretic interactions while simultaneously introducing anisotropic hydrodynamic resistance. Phoretic flows can interact with large-scale advection and provide mechanisms for microscale manipulation of colloidal particles, including size-dependent transport~\cite{Shin2015}, focusing~\cite{Ault2018}, band formation~\cite{Staffeld1989}, and sorting~\cite{Abecassis2008,Palacci2010}. A detailed understanding of the flow structures generated in confined geometries is therefore essential for predicting and controlling phoretic transport in microfluidic and porous environments.

Narrow channels, pores, and cavities present in many microfluidic and engineered systems provide particularly promising settings for such applications. Dead-end pores and confined cavities, for example, have been investigated as potential mechanisms for particle capture and removal~\cite{Wilson2020,Battat2019,Akdeniz2023,Alipour24,Li2026}, where concentration differences between the bulk fluid and confined regions can drive particle entrainment, trapping, and filtration in porous systems~\cite{Jotkar2024,Doan2021,Chu2021,Sambamoorthy2023,Teng2023,Somasundar2023,Sambamoorthy2025,Alessio2021,Kar2015}. Although such mechanisms and device concepts are actively explored through numerical simulations~\cite{Migacz2024,Bhattacharyya2023,Visan2024}, analytical descriptions of the coupled chemical and hydrodynamic fields in representative confined geometries remain essential. Such solutions provide fundamental insight, efficient numerical benchmarks, and reliable validation cases for continuum simulations.

Wedge geometries provide a particularly rich example of confined Stokes flows. Classical analyses date back to the study of Moffatt eddies in two-dimensional corners subject to surface-driven forcing~\cite{Moffatt_1964_JFM_18_1,Moffatt_1964_AMS_2_365} and to Taylor's scraper problem~\cite{taylor1962}. Although initially introduced as idealised mechanical forcing mechanisms and later realised experimentally~\cite{Taneda1979}, equivalent surface-driven flows can be generated through phoretic mechanisms, such as catalytic activity or localised heating, which provide experimentally accessible sources of slip. Subsequent studies have examined wedge and corner flows in the presence of free surfaces~\cite{liujoseph1977}, electrohydrodynamic forcing~\cite{He2022}, and three-dimensional conical geometries where vortical structures emerge~\cite{Shankar2005}. Similar localised recirculating flows have also been observed in driven cavity flows~\cite{Polychronopoulos2018,Biswas2018} and geophysical systems such as subglacial ice flows~\cite{Meyer2017}. 

From a mathematical perspective, wedge-shaped boundary-value problems are naturally treated using integral transform methods. In two-dimensional geometries, the Stokes equations can be reduced to a biharmonic equation for the stream function, allowing the Mellin transform to be employed effectively~\cite{TRANTER1948,Moffatt_1964_JFM_18_1,Moffatt_1964_AMS_2_365,Martin2017,Nowak_Lisicki_2026}. In fully three-dimensional Stokes flows, however, a scalar stream-function formulation is unavailable, requiring an alternative approach. Since solutions of the Stokes equations can be expressed in terms of harmonic functions, it is natural to represent the flow field using harmonic potentials within the Papkovich--Neuber formulation~\cite{tran1982general,Papkovich1932,Neuber1934}. The Fourier--Kontorovich--Lebedev (FKL) transform provides a particularly suitable framework for wedge geometries~\cite{kontorovich1938one,lebedev1949,erdelyi1953higher}. Specifically, Fourier transformation along the axial direction and Kontorovich--Lebedev transformation in the radial coordinate separate the corresponding variables while retaining the angular dependence, providing the basis of the approach adopted here.
The FKL transform has been successfully employed to address a range of classical Stokes-flow problems and their interactions with confining wedge boundaries~\cite{sano1978effect,Dauparas2018,daddi2026hydrodynamic}. For certain simplified configurations, the method of images has also been applied to related problems~\cite{Sprenger2023}.

The FKL representation yields compact spectral solutions for both reflecting and absorbing boundaries. For commensurate wedge angles, the concentration field can be reduced to finite image-like sums or closed expressions involving elementary and elliptic functions. The associated slip distribution generates a fully three-dimensional Stokes flow, represented through harmonic potentials with explicit spectral coefficients. Unlike the concentration field, the velocity solution generally requires numerical evaluation of spectral integrals because the hydrodynamic formulation introduces non-trivial spectral couplings and denominators~\cite{tran1982general,daddi2025proc}. A recent review~\cite{daddi2026spectral} summarises analytical approaches to wedge and corner flow problems.

The main contribution of this work is the development of a Green's-function formulation for three-dimensional phoretic flows in wedge geometries. We first solve the scalar diffusion problem and determine, in the low-Péclet-number regime, the resulting phoretic slip distribution. We then employ Papkovich--Neuber potentials to construct the induced Stokes flow while preserving the harmonic spectral structure of the concentration solution.
Beyond providing exact analytical solutions to a canonical transport problem, the resulting framework offers a versatile foundation for modelling chemically driven flows in confined microfluidic geometries, porous media, and patterned surfaces, while also serving as a benchmark for numerical methods.

The paper is organised as follows. In Sec.~\ref{sec:model}, we introduce the physical formulation of phoretic flows and the mathematical framework, including the relevant properties of the FKL transform. Section~\ref{sec:framework} presents the structure of the concentration and flow solutions and the strategy based on Papkovich--Neuber potentials. We derive and analyse the Green's function for Neumann--Neumann boundary conditions in Sec.~\ref{sec:neumann}, followed by the corresponding Neumann--Dirichlet solution in Sec.~\ref{sec:neumann-dirichlet}. Finally, we summarise our results and discuss future directions in Sec.~\ref{sec:conclusions}. The Appendix provides a detailed derivation of the integral identity used in constructing the concentration solution.

\section{Mathematical model}\label{sec:model}

\subsection{Phoretic flow}

\begin{figure}
    \centering    \includegraphics[width=0.5\linewidth]{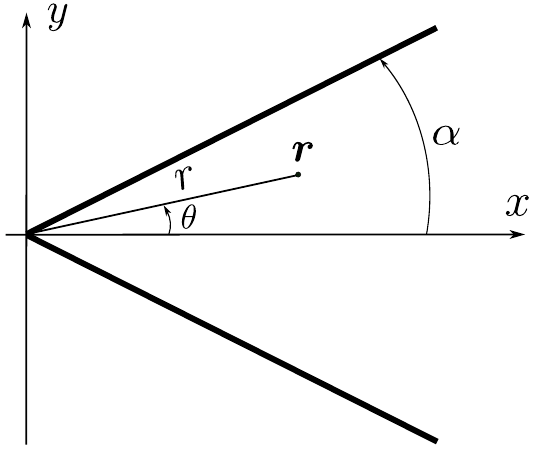}
    \caption{Schematic of a fluid confined within a wedge-shaped domain bounded by two planar walls. The geometry is described in cylindrical coordinates $(r,\theta,z)$, where $r$ denotes the radial distance from the wedge apex, $\theta$ is the polar angle measured from the centreline between the walls, and $z$ corresponds to the axial direction along the wedge edge.
    Through the surface activity of one or both boundaries, a phoretic flow is generated via tangential slip driven by gradients in solute concentration along the walls.
}
    \label{fig:placeholder}
\end{figure}

The continuum description of diffusiophoretic and diffusioosmotic transport was established in the pioneering works of Refs.~\cite{golestanian2007,julicher2009,sabass2012}.
In this article, we consider a three-dimensional wedge geometry bounded by two semi-infinite planar walls, denoted by $\mathcal{S}$, which intersect along a common edge and enclose an opening angle of $2\alpha$.
We restrict our attention to non-re-entrant wedge geometries, for which the semi-opening angle satisfies $\alpha\in(0,\pi/2]$.
We formulate the problem in cylindrical coordinates $(r,\theta,z)$, with $r$ measuring the distance from the wedge apex, $\theta$ defining the polar angle relative to the centreline between the walls, and $z$ denoting the axial coordinate along the wedge edge. The surface activity of one or both boundaries drives a phoretic flow by generating tangential slip velocities through solute concentration gradients established along the walls; see Fig.~\ref{fig:placeholder} for an illustration of the system setup.

In this work, we restrict the wedge half-angle to $\alpha\in(0,\pi/2]$. The complementary case of a salient wedge, $\alpha>\pi/2$, requires a separate treatment of the singular behaviour near the wedge edge, as discussed, for example, in the classical analysis of Stokes flow near a salient wedge~\cite{kim1983effect}.

The domain is occupied by a Newtonian fluid of density $\varrho$ and dynamic viscosity $\eta$, containing a dispersed solute with molecular diffusion coefficient $D$.
In the following, $c$ represents the number density of the solute molecules whose concentration gradients drive the phoretic flow.
Chemical activity at the walls drives the release or absorption of solute molecules, which we assume occurs at a prescribed rate. We quantify the strength of this process through the surface activity~$\mathcal{A}$, which characterises the local reaction rate at the boundaries. The phoretic mobility $\mathcal{M}$ characterises the coupling between tangential solute concentration gradients and the resulting surface slip velocity~\cite{golestanian_les_houches}.
In general, both the surface activity $\mathcal{A}$ and the phoretic mobility $\mathcal{M}$ may vary spatially along the surface. Consequently, the resulting solute concentration and phoretic flow can exhibit a fully three-dimensional dependence on $(r,\theta,z)$.

We denote by $|\mathcal{A}|$ and $|\mathcal{M}|$ the maximum magnitudes of the chemical activity and phoretic mobility prescribed on the surfaces, respectively. 
To nondimensionalise the governing equations and identify the relevant control parameters, we introduce a characteristic length scale $\mathcal{L}$.
The concentration scale is then defined as $\mathcal{C}=|\mathcal{A}|\mathcal{L}/D$, which follows from the magnitude of the solute gradients generated at the surfaces. The characteristic magnitude of the induced surface flow is given by $\mathcal{V}=|\mathcal{A}\mathcal{M}|/D$, while the corresponding pressure scale follows from the viscous stress balance as $\mathcal{P}=\eta\mathcal{V}/\mathcal{L}$.

The transport physics is characterised by two dimensionless parameters: the Péclet number,
\begin{equation}
\text{Pe} = \frac{\mathcal{V}\mathcal{L}}{D} \, ,
\end{equation}
which measures the relative importance of solute advection compared with molecular diffusion, and the Reynolds number,
\begin{equation}
\text{Re} = \frac{\varrho \mathcal{V}\mathcal{L}}{\eta} \, ,
\end{equation}
which characterises the relative importance of inertial and viscous effects in the flow. We focus on the regime relevant to microfluidic applications, where solute transport is dominated by diffusion and the fluid motion is governed by viscous stresses. Accordingly, we consider the limit $\text{Pe}\ll 1$ and $\text{Re}\ll 1$.
This implies that the solute concentration field satisfies the steady-state Laplace equation. In nondimensional variables, this equation takes the form 
\begin{equation} \label{eq:laplace}
    \nabla^2 c = 0 \, ,
\end{equation}
and the flow field satisfies the incompressible Stokes equations~\cite{kim05}
\begin{equation}
    \nabla^2 \bm{v} = \nabla p\, , \qquad
    \nabla\cdot\bm{v} =0 \, .
    \label{eq:Stokes}
\end{equation}

The local solute flux prescribed on the boundary $\mathcal{S}$ takes the form
\begin{equation}\label{eq:fixedflux}
\bm{n}\cdot\nabla {c} =-A\quad \textrm{      on   } \mathcal{S} \, ,
\end{equation}
with $\bm{n}$ denoting the outward unit normal vector to the boundary.
In addition, $A=\mathcal{A}/|\mathcal{A}|$ denotes the dimensionless activity, with regions satisfying $A>0$ acting as solute sources and those with $A<0$ acting as solute sinks. The generation of solute gradients, combined with short-range interactions between solute molecules and the boundary, induces motion of the surrounding fluid through phoretic effects~\cite{anderson1989}. Since the characteristic interaction length is small compared with the characteristic dimensions of the system, the resulting near-wall motion can be represented by an effective slip velocity boundary condition for the bulk Stokes flow within the cavity~\cite{michelin2014}. Within this framework, the surface slip velocity is proportional to the tangential gradient of the solute concentration 
\begin{equation}\label{eq:slip_def}
{\bm{v}} =M (\boldsymbol{1}-\bm{n}\bm{n})\cdot\nabla {c}\quad \textrm{      on}\,\, \mathcal{S} \, ,
\end{equation}
where $M=\mathcal{M}/|\mathcal{M}|$.

In this limit, the coupling between the flow field and the solute concentration is unidirectional: the solution of the stationary diffusion problem determines the boundary condition driving the flow, while the flow does not influence the concentration field. We further note that the problem is invariant under an arbitrary shift in the absolute value of the concentration, since both Eq.~\eqref{eq:laplace} and the boundary condition, Eq.~\eqref{eq:fixedflux}, depend only on concentration gradients. It is therefore convenient to define the concentration relative to the far-field equilibrium concentration, which removes the irrelevant additive constant and ensures that the concentration field represents only the deviations generated by the surface activity.

Taken together, Eqs.~\eqref{eq:laplace}--\eqref{eq:slip_def} constitute the full problem of flow determination for given chemical properties of the surface. In the following, we will focus on solving them for given physical contexts and boundary conditions on the walls.

\subsection{Fourier-Kontorovich-Lebedev transform}

To determine the solute concentration and fluid velocity fields in the wedge-shaped domain, we employ the Fourier–Kontorovich–Lebedev transform.
The method is particularly well suited to boundary-value problems in wedge geometries by mapping the axial and radial coordinates onto the spectral variables $k$ and $\nu$, respectively. 
The FKL transform can be viewed as analogous to the more familiar Fourier–Hankel transform used in cylindrical geometries. Here, the Fourier transform is applied along the translationally invariant $z$ direction, while the Kontorovich–Lebedev transform is applied in the radial coordinate $r$. The latter is particularly well suited to the wedge geometry, as it naturally leads to modified Bessel functions of imaginary order, $\K_{i\nu}$, and facilitates the imposition of boundary conditions on the wedge surfaces.

While the Fourier transform is widely used, the Kontorovich–Lebedev transform is less familiar; it was originally introduced by the renowned Russian mathematicians  Kontorovich and Lebedev to treat specific classes of boundary-value problems and later developed in a series of foundational works, with further details given in Erdélyi \textit{et~al.}~\cite[p.~75]{erdelyi1953higher}.
This transform has since found applications in a variety of wedge-related problems, including electromagnetic scattering and diffraction~\cite{rawlins1999diffraction, antipov2002diffraction}, elasticity~\cite{daddi2025proc, daddi2025jelasticity}, and fluid dynamics~\cite{sano1978effect, daddi2026hydrodynamic, daddi2026selfdiffusio}.
In recent years, Yakubovich and his collaborators have made several important contributions to the foundations of the KL transform~\cite{yakubovich2003kontorovich, yakubovich2008progress, yakubovich2012index, rodrigues2013convolution, loureiro2013central, yakubovich2024index}.

In the following, FKL-transformed functions are denoted by a tilde.
We define the forward FKL transform as
\begin{equation}
    \widetilde{f}(\nu,\theta,k) :=
    \mathscr{T}_{i\nu} \{f\} = 
    \int_{-\infty}^\infty \mathrm{d}z \, e^{ikz} \int_0^\infty f(r,\theta, z) \,\K_{i\nu} \left( |k|r\right) \, r^{-1} \, \mathrm{d} r \, , 
\end{equation}
where $\K_{i\nu}(|k|r)$ denotes the modified Bessel function of the second kind of imaginary order $i\nu$.
Note that the polar angle is unaffected by the transform.
The inverse transform is obtained as
\begin{equation}
   f(r,\theta, z) = \mathscr{T}_{i\nu}^{-1} \{f\}
   = \frac{1}{\pi^3} 
   \int_{-\infty}^\infty \mathrm{d}k\, 
   e^{-ikz} 
   \int_0^\infty \widetilde{f}(\nu,\theta, k) \, \K_{i\nu} (|k|r) \sh (\pi\nu) \, \nu \, \mathrm{d} \nu \, .
   \label{eq:inv_FKL_def}
\end{equation} 
Throughout this work, we use the abbreviations sh, ch, th, cth, sch, csh, and ach to denote the hyperbolic sine, cosine, tangent, cotangent, secant, cosecant, and inverse hyperbolic cosine functions, respectively.

In many applied problems, it is considerably simpler to perform the inverse Fourier transform, so that the solution can be written as a single infinite integral over the radial wavenumber $\nu$.
We note that, for $\xi>0$, the modified Bessel function satisfies the relation $\overline{\K_{\mu}(\xi)}=\K_{\overline{\mu}}(\xi)$ for $\mu\in\mathbb{C}$. Therefore, for $\mu=i\nu$, with $\nu\in\mathbb{R}$, we have $\overline{\K_{i\nu}(\xi)}=\K_{-i\nu}(\xi)=\K_{i\nu}(\xi)$, since the modified Bessel function is an even function of its order. Hence, $\K_{i\nu}(\xi)$ is real-valued in this case.

By applying the FKL transform to the governing equations, the problem reduces to an ordinary differential equation in the polar angle $\theta$, which can be solved straightforwardly. In FKL space, the Laplace equation reduces to solving a second-order linear differential equation of the form
\begin{equation}
     \left( \frac{\partial^2}{\partial\theta^2}-\nu^2 \right) \widetilde{f} = 0 \, .     \label{eq:laplace_FKL}
\end{equation}

\section{General framework}\label{sec:framework}

\subsection{Concentration field}

Since the concentration field is harmonic, its general solution in FKL space is obtained from the solution of Eq.~\eqref{eq:laplace_FKL} and can be written as 
\begin{equation}
     \widetilde{c}(\nu,\theta,k) = g(\nu,k) \sh (\theta\nu) + g^\dagger(\nu,k) \ch (\theta\nu) \, ,
     \label{eq:FKL_general_solution}
\end{equation}
where $g$ and $g^\dagger$ are unknown spectral functions that are determined from the prescribed boundary conditions.

The boundary conditions on the wedge faces are formulated in two standard combinations. In the Neumann–Neumann case, the normal derivatives of the concentration are prescribed on both faces, corresponding to specified surface activity through the normal gradient of concentration on each boundary. In the mixed Neumann-Dirichlet case, the normal derivative is prescribed on one wedge face while concentration is prescribed on the other.
In all cases, homogeneous limits correspond to zero concentration for Dirichlet conditions or zero normal flux (no activity) for Neumann conditions.

From a physical perspective, the boundary conditions for the solute concentration are determined by the reaction kinetics at the solid–fluid interface \cite{michelin2014,sabass12}. A Neumann condition corresponds to a prescribed surface flux and is appropriate when the reaction rate is effectively independent of the local solute concentration, as in the fixed-flux approximation of Eq.~\eqref{eq:fixedflux}, commonly employed for chemically active surfaces \cite{michelin2014,golestanian07}. More generally, finite-rate reaction kinetics lead to a Robin boundary condition, relating the normal solute flux to the local surface concentration. The relative importance of reaction kinetics and bulk diffusion is characterised by the Damköhler number \cite{michelin2014,sabass12}. In the limit of sufficiently fast reactions or rapid exchange with a chemical reservoir, the surface concentration may instead become effectively fixed, corresponding to a Dirichlet condition.

Dirichlet–Dirichlet conditions therefore have a physically admissible interpretation, for example when both wedge faces act as ideal sinks or are maintained at prescribed concentrations. However, this configuration differs from the activity-driven phoretic pumping considered here. In particular, spatially uniform concentrations imposed on both walls may sustain diffusive transport across the wedge but do not generate the tangential concentration gradients required to drive diffusioosmotic slip. We therefore focus on Neumann–Neumann and mixed Neumann–Dirichlet conditions, which describe flow generated by a localised chemically active patch in the presence of either a reflecting or a concentration-controlled opposing boundary.

In the two cases considered, with corresponding fields denoted by subscripts N and M, respectively, the unknown functions in Eq.~\eqref{eq:FKL_general_solution} are determined from the boundary conditions imposed on each wedge face, written as
\begin{enumerate}[label=(\alph*)]
    \item For Neumann–Neumann boundary conditions:
    \begin{equation}
         -\frac{1}{r}\frac{\partial c}{\partial \theta}\bigg|_{\theta=\pm\alpha} = A_\pm(r,z) \, ,
    \end{equation}
where $A_\pm$ denote the surface activities imposed on each wedge face.
Defining the abbreviation $B_\pm (r,z) = r A_\pm (r,z)$, the solution for the unknown coefficients in FKL space is obtained as 
\begin{subequations}
    \begin{align}
    g(\nu, k) &= -\frac{1}{2\nu} \left( \widetilde{B}_+ + \widetilde{B}_- \right) \sch(\alpha\nu)\, , \\
    g^\dagger(\nu, k) &= \phantom{+} \frac{1}{2\nu} \left( \widetilde{B}_- - \widetilde{B}_+ \right) \csh(\alpha\nu) \, .
\end{align}
\end{subequations}
The resulting concentration field is given by
    \begin{equation}
        \widetilde{c}_\mathrm{N} (\nu,\theta,k)
        = \frac{1}{\nu}
        \left( 
        \widetilde{B}_-\ch \left( (\alpha-\theta)\nu\right)
        -\widetilde{B}_+ \ch \left( (\alpha+\theta)\nu\right)
        \right) \csh(2\alpha\nu) \, .
        \label{eq:concentration_neu_neu}
    \end{equation}    
    \item For mixed Dirichlet–Neumann boundary conditions:
    \begin{equation}
    -\frac{1}{r}\frac{\partial c}{\partial \theta}\bigg|_{\theta=-\alpha} = A_-(r,z)  \, ,
    \qquad 
    c(r,\theta=\alpha,z) = C_+(r,z)\, ,
\end{equation}
the spectral functions are obtained as
\begin{subequations}
    \begin{align}
    g(\nu, k) &= \left( C_+\sh(\alpha\nu) - \frac{B_-}{\nu}\, \ch(\alpha\nu) \right)\sch(2\alpha\nu)\, , \\
    g^\dagger(\nu, k) &= \left( C_+\ch(\alpha\nu) + \frac{B_-}{\nu}\, \sh(\alpha\nu) \right)\sch(2\alpha\nu) .
\end{align}
\end{subequations}
This yields the concentration field in FKL space as
\begin{equation}
    \widetilde{c}_\mathrm{M} (\nu,\theta,k) = 
     \left( \widetilde{C}_+ \ch \left( (\alpha+\theta)\nu\right)
    + \frac{ \widetilde{B}_- }{\nu}\, 
    \sh \left( (\alpha-\theta)\nu \right)
    \right) \sch (2\alpha\nu) \, .
    \label{eq:concentration_mixed}
\end{equation}

\end{enumerate}

In the present work, we consider a configuration in which the surface-driven flow is induced by a point active patch on the wedge boundary located at $\theta=-\alpha$. 
This is implemented mathematically by introducing a Dirac delta function to represent the flux at this boundary.
We examine two cases for the opposing boundary at $\theta=\alpha$, namely zero-flux and zero-concentration conditions.
Exploiting the linearity of the governing equations, the solution for an arbitrary surface activity can be constructed directly from this fundamental solution, also termed the Green’s function.

The solution for the concentration and flow velocity fields in real space can be obtained via an inverse FKL transform involving a double integration with respect to the wavenumbers $k$ and $\nu$. While the integration with respect to $k$ can be carried out using tabulated integrals involving products of modified Bessel functions and trigonometric functions, one is left with more delicate integrals over $\nu$.
Analytical progress is possible only in the case of the concentration field and for commensurate angles, and even then it remains non-trivial. For the velocity field, the situation is significantly more challenging. This is due to the fact that the relevant poles are determined by transcendental equations that do not admit closed-form solutions, which precludes a direct application of the residue theorem.
As a result, the flow field solution will be obtained using numerical integration.

As will be shown in the sequel, the Green's function for wedge problems can generally be expressed in the form $g(\nu,k)=G\,\mathrm{K}_{i\nu}(|k|\rho)$ and $g^\dagger(\nu,k)=G^\dagger\,\mathrm{K}_{i\nu}(|k|\rho)$, where $\rho$ denotes the radial distance locating the singularity relative to the wedge apex, and the coefficients $G$ and $G^\dagger$ depend solely on the transform variable $\nu$ and the wedge semi-opening angle $\alpha$. Accordingly, the real-space concentration field is obtained via the inverse FKL transform given by Eq.~\eqref{eq:inv_FKL_def}, yielding
\begin{equation}
     c(r,\theta,z) = \frac{1}{2\pi \sqrt{\rho r}}
    \int_0^\infty 
    \left( G \sh(\theta\nu) + G^\dagger \ch(\theta\nu) \right)
   \th(\pi\nu) \, \mathrm{P}_{i\nu-\frac{1} {2}}(\mu) \,  \nu \,\mathrm{d}\nu \, , 
    \label{eq:c_real_space_def}
\end{equation}
where $\mathrm{P}_n$ denotes the Legendre function of the first kind of degree $n$~\cite{abramowitz72}, with argument $\mu=(r^2+\rho^2+z^2)/(2\rho r)$, where $\mu\geq 1$.
Here, we have employed the following known integral.
\begin{equation}
    \frac{4}{\pi^2} \int_0^\infty  \K_{i\nu}(k\rho) \, \K_{i\nu}(kr)\cos(kz) \, \mathrm{d}k 
    =
     \frac{1}{\sqrt{\rho r}} \, 
     \sch(\pi\nu)
     \, \mathrm{P}_{i\nu-\frac{1}{2}} (\mu)  \, ;
     \label{eq:Kinu_analytical_result}
\end{equation}
see, for example, Gradshteyn and Ryzhik~\cite[p.~719, Eq.~6.672.3]{gradshteyn2014table} or Prudnikov \textit{et~al.}~\cite[p.~390, Eq.~2.16.36.2]{prudnikov1992integrals}. The Legendre function of the first kind can be expressed in an integral form as
\begin{equation}
    \mathrm{P}_{i\nu-\frac12} (\mu) = 
    \frac{\sqrt{2}}{\pi} \, \cth(\pi \nu) \int_{\operatorname{ach}\mu}^\infty 
\frac{\sin (\nu t) \, \mathrm{d}t}{\sqrt{\ch t-\mu}} \, , 
\label{eq:LegendreP_int}
\end{equation}
with $\operatorname{ach} = \ch^{-1}$ denoting the inverse hyperbolic cosine function.
The integral form given in Eq.~\eqref{eq:LegendreP_int} enables further analytical progress by allowing the interchange of integrals and the subsequent evaluation via double integration, as detailed in the present work.
Using this integral representation has the advantage that $\cth$ cancels with $\th$ in Eq.~\eqref{eq:c_real_space_def}, leading to a simplified expression, albeit with a double integral.

It is worth noting that the FKL representation remains applicable in the vicinity of the wedge apex, where $r\to0$. 
The transform representation is well suited to handling the corner singularity. In fact, from Eq.~\eqref{eq:Kinu_analytical_result} and the Cauchy–Schwarz inequality, we obtain
\begin{align}
    \frac{\sch(\pi\nu)}{\sqrt{\rho r}}\left|P_{i\nu-1/2}(\mu)\right|
&\leq \frac{4}{\pi^2}\int_0^\infty
\left|\mathrm{K}_{i\nu}(k\rho)\mathrm{K}_{i\nu}(kr)\cos(kz)\right|\,\mathrm{d}k
    \leq \frac{4}{\pi^2}\int_0^\infty \mathrm{K}_0(k\rho)\mathrm{K}_0(kr)\,\mathrm{d}k \notag \\
&\leq \frac{4}{\pi^2}
\left(\int_0^\infty \mathrm{K}_0^2(k\rho)\,\mathrm{d}k\right)^{1/2}
\left(\int_0^\infty \mathrm{K}_0^2(kr)\,\mathrm{d}k\right)^{1/2} 
= {4\over \pi^2 \sqrt {\rho r}} \int_0^\infty  \mathrm{K}^2_{0}(k) \, \mathrm{d}k
=\frac{1}{\sqrt{\rho r}} \, , 
\end{align}
where we have used formula 2.16.33.2 in \cite[Vol.~2]{yakubovich2024index}. Hence,
\begin{equation}
    \left|P_{i\nu-1/2}(\mu)\right|\leq \ch(\pi\nu) \, ,
\end{equation}
which provides a uniform bound for the Legendre function. Consequently, the right-hand side of (3.9) has, at most, an integrable singularity of order $O\left(r^{-1/2}\right)$ as the corner is approached. Thus, although the solution may become singular at the corner, the FKL transform representation remains well defined and provides a controlled, integrable description of the singular behaviour.

\subsection{Hydrodynamic flow field}

For simplicity, we assume that the phoretic mobility is uniform over the surface of the wedges. This significantly simplifies the mathematical analysis, although cases with spatially varying surface mobility can also be treated within the same framework. The slip velocity is obtained from the tangential gradient of the concentration field evaluated at the surface, which in the cylindrical coordinate system has radial and axial components given by
\begin{equation}
    {v_r^\mathrm{S}}_\pm = \left. M\, \frac{\partial c}{\partial r} \right|_{\theta=\pm\alpha} \, , \qquad
    {v_z^\mathrm{S}}_\pm = \left. M\, \frac{\partial c}{\partial z} \right|_{\theta=\pm\alpha} \, .
    \label{eq:slip_velo_def}
\end{equation}

Equations~\eqref{eq:slip_velo_def} serve as boundary conditions for the hydrodynamic problem describing a surface-driven flow induced by phoretic slip at the interface.

The general solution of the Stokes equations~\eqref{eq:Stokes} can be written in terms of the Papkovich–Neuber representation as~\cite{Papkovich1932,Neuber1934, tran1982general}
\begin{equation}
    \bm{v} (\bm{r}) = \bm{\nabla} \bigl( \bm{r} \cdot \boldsymbol{\upPhi} (\bm{r}) + \rho \, \upPhi_w (\bm{r}) \bigr) - 2 \, \boldsymbol{\upPhi} (\bm{r}) \, , 
    \qquad
    p(\bm{r}) = 2 \, \boldsymbol{\nabla} \cdot \upPhi (\bm{r}) \,  ,
    \label{eq:Papkovich_Neuber}
\end{equation}
where $\boldsymbol{\upPhi} (\bm{r})$ is composed of the Cartesian components $\upPhi_x$, $\upPhi_y$, and $\upPhi_z$.

In what follows, the explicit dependence of $\bm{v}$, $p$, and $\boldsymbol{\upPhi}$ on $\bm{r}$ is omitted. Each component $\upPhi_j$ is a harmonic function that satisfies Laplace’s equation.
In cylindrical coordinates, the velocity components take the form
\begin{equation}
   v_r = \frac{\partial \upPi}{\partial r} - 2\upPhi_r \,, \qquad
v_\theta = \frac{1}{r} \frac{\partial \upPi}{\partial \theta} - 2\upPhi_\theta \,, \qquad
v_z = \frac{\partial \upPi}{\partial z} - 2\upPhi_z \, , 
   \label{eq:velo_cylindrical_coord}
\end{equation}
where we have defined the combination of components
\begin{equation}
    \upPi = r\upPhi_r + z\upPhi_z+ \rho \, \upPhi_w \, .
    \label{eq:Pi_fun}
\end{equation}
In addition, the \emph{non-harmonic} functions $\upPhi_r = \upPhi_x \cos\theta + \upPhi_y \sin\theta$ and $\upPhi_\theta =  - \upPhi_x \sin\theta + \upPhi_y \cos\theta$ denote the radial and azimuthal components of $\boldsymbol{\upPhi}$, respectively.

Being harmonic functions, the general solutions for $\upPhi_j$, $j \in \{x,y,z,w\}$, have a form analogous to that given in Eq.~\eqref{eq:FKL_general_solution} in FKL space and can be written as
\begin{equation}
    \widetilde{\upPhi}_j (\nu,\theta,k) = 
    F_j(\nu,k) \sh(\theta\nu) + F_j^\dagger(\nu,k) \ch(\theta \nu) \, , \qquad 
    j \in \{x,y,z,w\} \, .
    \label{eq:general_solution}
\end{equation}

We have a total of eight unknown coefficients to determine from the velocity boundary conditions imposed on the wedge surfaces, comprising impermeability conditions requiring zero normal velocity at the boundaries together with prescribed tangential velocities given by the slip conditions.
Specifically
\begin{equation}
   \left. \frac{\partial \upPsi}{\partial r} - 2\upPhi_r \right|_{\theta=\pm\alpha} = 0 \,, \qquad
    \left. \frac{1}{r} \frac{\partial \upPi}{\partial \theta} - 2\upPhi_\theta \right|_{\theta=\pm\alpha} = 0 \,, \qquad
    \left. \frac{\partial \upPsi}{\partial z} - 2\upPhi_z \right|_{\theta=\pm\alpha} = 0 \, ,
   \label{eq:BCs}
\end{equation}
where we have defined
\begin{equation}
    \upPsi = \upPi - Mc \, .
\end{equation}

Since these boundary conditions yield six equations when imposed on both boundaries, we retain two remaining degrees of freedom in the choice of coefficients. A convenient choice that leads to a simplified mathematical treatment is to assume that the function~$\upPsi$ vanishes on the boundaries.
In this way, the problem becomes well posed, and the hydrodynamic flow field can be fully determined from the prescribed slip profile on the wedge surfaces.
We note that this condition implies that the tangential derivatives of $\upPsi$, namely those with respect to $r$ and $z$, also vanish on the wedge surfaces. Consequently, both $\upPhi_r$ and $\upPhi_z$ vanish on the wedge surfaces as well.

As $\upPhi_z$ is a harmonic function that vanishes at $\theta=\pm\alpha$, it follows from the general solution given by Eq.~\eqref{eq:FKL_general_solution} that $\upPhi_z=0$ everywhere. 
This does not apply to $\upPhi_r$, since it is not a harmonic function; rather, it is a field that must be determined as part of the solution.
Furthermore, since $c$ is a harmonic function whose expression can be written in FKL space in the form given by Eq.~\eqref{eq:FKL_general_solution}, the condition $\upPsi=0$ at $\theta=\pm\alpha$ implies that $\rho\, \upPhi_w=Mc$.

We are therefore left with determining the four coefficients defining the functions $\upPhi_r$ and $\upPhi_\theta$, which are themselves not harmonic. These coefficients can be obtained by enforcing the remaining boundary conditions, which in FKL space can be written as
\begin{equation}
    \left. \widetilde{\upPhi}_r \right|_{\theta=\pm\alpha} = 0 \, , \qquad
    -2\upPhi_\theta  + \left. \frac{\partial}{\partial\theta} \,\mathscr{T}_{i\nu}
    \left\{ \upPhi_r + \frac{1}{r}\, \upPhi_w \right\} \right|_{\theta=\pm\alpha} = 0 \, .
    \label{eq:four-eqns-BCs}
\end{equation}

We note that the FKL transform of a function divided by $r$ introduces nonlocal effects that alter the order of the index transform. This property, derived in earlier work~\cite{daddi2026spectral}, is given by
\begin{equation}
    \mathscr{T}_{i\nu} \left\{ \frac{f}{ r} \right\}
    = 
    \frac{|k|}{2i\nu} 
    \big(  \mathscr{T}_{i\nu+1} \{f\}
    -  \mathscr{T}_{i\nu-1} \{f\}
    \big) .
    \label{eq:FKL_f_over_r}
\end{equation}

For the presentation of the solution for the induced phoretic flow field, we express the coefficients associated with the harmonic functions in the form
\begin{equation}
    F_j = \frac{M}{\rho} \, \Lambda_j \, \K_{i\nu} \left( |k|\rho\right) \, ,
    \qquad
    F_j^\dagger = \frac{M}{\rho} \,\Lambda_j^\dagger  \, \K_{i\nu} \left( |k|\rho\right) \, , \qquad j\in\{w,x,y\} \, , 
    \label{eq:coef_form}
\end{equation}
where $\Lambda_j$ and $\Lambda_j^\dagger$ are functions of $\nu$ and $\alpha$, and may also depend on the operator $\rho\, \partial / \partial \rho$.
The purpose of this representation is to isolate the explicit dependence of the coefficients $F_j$ and $F_j^\dagger$ from $k$, which facilitates the inverse Fourier transform that is performed prior to the inverse KL transform.

The solution for the velocity field in real space can be written in integral form using the FKL-space representation given in Eq.~\eqref{eq:general_solution}, together with the expressions for the coefficients in
Eq.~\eqref{eq:coef_form} as
\begin{equation}
   \upPhi_j (r, \theta, z) = \frac{M}{2\pi \rho \sqrt{\rho r}}
   \int_0^\infty 
   \left(  \Lambda_j  \sh(\theta\nu) + \Lambda_j^\dagger  \ch(\theta\nu)  \right) \th(\pi\nu) 
   \, \mathrm{P}_{i\nu-\frac{1}{2}}(\mu) \,
   \nu \, \mathrm{d}\nu \, , 
    \label{eq:phi_real_space}
\end{equation}
for $j \in \{x,y,w\}$.
Except for $\upPhi_w$, whose expression is directly obtained as $\upPhi_w = Mc/\rho$, a closed-form expression for the harmonic functions $\upPhi_x$ and $\upPhi_y$ is considerably more delicate, even in the commensurate-angle case, since the poles cannot be determined explicitly in closed form; see Eq.~\eqref{eq:delta_pm}. As a consequence, the velocity field can only be computed via numerical integration.

In the following, we derive the Green's functions for the two problem configurations considered in this study, beginning with the concentration field. Exact analytical solutions are obtained for wedges with commensurate opening angles.

\section{Green's function for Neumann–Neumann boundary conditions}\label{sec:neumann}

We consider Neumann–Neumann boundary conditions, where the phoretic flow is driven by a point activity patch located on the lower wedge at $(r,\theta,z) = (\rho,-\alpha,0)$, while no-flux conditions are imposed on the upper wedge. We require that
\begin{subequations}
    \begin{align}
    -\frac{1}{r}\, \frac{\partial c}{\partial \theta}
    &= \delta(r-\rho) \, \delta(z) \,\,\quad \text{for}\quad \theta=-\alpha \, , \\
    \frac{\partial c}{\partial \theta}
    &= 0 \hspace{6em} \, \text{for}\quad \theta=\alpha \, .
\end{align}
\end{subequations}

\subsection{Solution in FKL space}

The problem corresponds to Neumann–Neumann boundary conditions with $A_-(r,z) =\delta(r-\rho)\,\delta(z)$ and $A_+(r,z)=0$, which yield the FKL transforms $\widetilde{B}_- = \K_{i\nu}(|k|\rho)$ and $\widetilde{B}_+=0$.
Substituting these expressions into Eq.~\eqref{eq:concentration_neu_neu} yields the induced concentration field in the form
\begin{equation}
    \widetilde{c} (\nu, \theta, k) = \frac{1}{\nu} \, \csh(2\alpha\nu) \ch \left( (\alpha-\theta)\nu\right) \K_{i\nu}(|k|\rho) \, .
    \label{eq:c_FKL_NN}
\end{equation}

The corresponding solution for the harmonic function $\upPhi_w$ is straightforward and given by $\upPhi_w = Mc/\rho$, from which it follows that
\begin{equation}
    \Lambda_w = -\frac{1}{2\nu}\, \sch(\alpha\nu) \sh(\pi\nu) \, , \qquad
    \Lambda_w^\dagger = \frac{1}{2\nu}\, \csh(\alpha\nu) \sh(\pi\nu) \, .
\end{equation}

The solution for $\upPhi_r$ and $\upPhi_\theta$ follows directly from solving the linear system of four equations given in Eqs.~\eqref{eq:four-eqns-BCs}. For the FKL transform of $\upPhi_w/r$, we use Eq.~\eqref{eq:FKL_f_over_r}. This yields expressions for the coefficients $F_j$ and $F_j^\dagger$, with $j \in \{x,y\}$.
We obtain
\begin{subequations}
    \begin{align}
    \Lambda_x &= \delta_+ \sin\alpha \ch(\alpha\nu)\, , \qquad\,
    \Lambda_x^\dagger = -\delta_- \sin\alpha \sh(\alpha\nu)\, , \\
    \Lambda_y &= \delta_- \cos\alpha \ch(\alpha\nu)\, , \qquad
    \Lambda_y^\dagger = -\delta_+ \cos\alpha \sh(\alpha\nu)\, , 
\end{align}
\end{subequations}
where
\begin{equation}
    \delta_\pm = 
    \frac{1}{\sh(2\alpha\nu)\pm\nu\sin(2\alpha)} \, .
    \label{eq:delta_pm}
\end{equation}

\subsection{Concentration field in real space}

From the expression for the concentration field in FKL space given by Eq.~\eqref{eq:c_FKL_NN}, the corresponding real-space solution is obtained via the inverse FKL transform in Eq.~\eqref{eq:c_real_space_def}, yielding
\begin{equation}
     c(r, \theta, z) = \frac{1}{2\pi \sqrt{\rho r}}
    \int_0^\infty 
    \csh (2\alpha\nu)  \ch \left( (\alpha-\theta)\nu \right) \th(\pi\nu)\, 
    \mathrm{P}_{i\nu-\frac{1}{2}}(\mu) \, \mathrm{d}\nu \, .
    \label{eq:concentration_NN}
\end{equation}

We now proceed to make analytical progress towards deriving an exact solution for the concentration field in real space.
Substituting the integral representation of the Legendre function given by Eq.~\eqref{eq:LegendreP_int} into Eq.~\eqref{eq:concentration_NN} yields the concentration field in the form of a double integral, namely,
\begin{equation}
    c(r, \theta, z) = \frac{1}{\pi^2 \sqrt{2\rho r}}
    \int_{\operatorname{ach}\mu}^\infty 
    \frac{\mathrm{d}t}{\sqrt{\ch t-\mu}}
    \int_0^\infty 
    \sin(\nu t)
    \csh(2\alpha\nu)
    \ch \left( (\alpha-\theta)\nu\right)
     \mathrm{d}\nu \, ,
\end{equation}
where we have interchanged the order of the two integrals provided that the integrand is absolutely integrable over the product domain, so that Fubini’s theorem applies.
Absolute convergence can readily be established by noting that $|\sin(\nu t)| \le \nu t$ and by examining the behavior of the remaining factors in the integration limits.

The integral over the radial wavenumber $\nu$ admits a closed-form expression, given by
\begin{equation}
    \int_0^\infty 
    \sin(\nu t)\ch \left( (\alpha-\theta)\nu\right)
    \csh(2\alpha\nu) \, \mathrm{d}\nu
    = \frac{\pi}{4\alpha}
    \frac{\sh  \frac{\pi t}{2\alpha} }{\ch  \frac{\pi t}{2\alpha}  + \sin  \frac{\pi\theta}{2\alpha} } \, ;
    \label{eq:Gradshteyn1}
\end{equation}
see, for instance, Gradshteyn and Ryzhik~\cite[p.~510, Eq.~3.981.8]{gradshteyn2014table}.

To the best of our knowledge, the resulting integral with respect to $t$ does not admit an analytic evaluation. Analytical progress can be made by assuming that $\alpha$ is a commensurate angle of the form $\alpha = \pi/q$, where $q$ is an integer. 
Depending on the parity of $q$, i.e.\ whether $q$ is even or odd, the corresponding formulas can be obtained for each case.
We further define $\theta_k=\theta-(4k-1)\,\pi/q$, where $k=1,\dots,n$ when $q=2n$, and $k=1,\dots,2n+1$ when $q=2n+1$.

We first consider the case $q=2n$. By making use of the integral identity given in Eq.~\eqref{eq:Gradshteyn1}, the concentration field can be reduced to a single improper integral, namely,
\begin{equation}
    c(r, \theta, z) = \frac{1}{2\pi^2 \sqrt{2\rho r}}    \int_{\operatorname{ach}\mu}^\infty 
    \frac{n \sh(nt)}{\ch(nt)+\sin(n\theta)}
    \frac{\mathrm{d}t}{\sqrt{\ch t-\mu}} \, .
    \label{eq:c_NN_even_tmp}
\end{equation}

This integral is convergent and admits a closed-form analytical expression. Since the derivation is somewhat involved, we defer the details to the Appendix, where we show that its evaluation yields
\begin{equation}
    c(r, \theta, z) = \frac{1}{2\pi \sqrt{2\rho r}}
    \sum_{k=1}^{n} \frac{1}{\sqrt{\mu - \cos \theta_k}}\, .
    \label{eq:c_commensurate_EVEN}
\end{equation}

The case $q=2n+1$ is slightly more delicate, but an analytical evaluation can nevertheless be obtained; the details are relegated to the Appendix. In this case, the concentration field can be written as a single improper integral by making use of Eq.~\eqref{eq:Gradshteyn1}, namely,
\begin{equation}
    c(r, \theta, z) = \frac{1}{2\pi^2 \sqrt{2\rho r}}
    \bigintsss_{\operatorname{ach}\mu}^\infty 
    \frac{ \left( n+\frac12 \right) \sh\left(\left(n+\frac{1}{2} \right)t\right)}{\ch\left(\left(n+\frac{1}{2} \right)t\right)+\sin\left(\left(n+\frac{1}{2}\right)\theta\right)}
    \frac{\mathrm{d}t}{\sqrt{\ch t-\mu}} \, .
    \label{eq:c_NN_odd_tmp}
\end{equation}
This is evaluated as
\begin{equation}
    c(r, \theta, z) = \frac{1}{2\pi^2 \sqrt{2\rho r}}
    \sum_{k=1}^{2n+1}
    \frac{1}{\sqrt{\mu-\cos\theta_k}}
    \operatorname{acos} \left( - \tfrac{1}{\beta}\, \cos \tfrac{\theta_k}{2} \right) ,
    \label{eq:c_commensurate_ODD}
\end{equation}
where we have introduced the abbreviation $\beta=\sqrt{(\mu+1)/2}$, which satisfies $\beta \ge 1$ since $\mu \ge 1$.
We note that, on the principal branch and for $x\in[-1,1]$, we use the identity $\operatorname{acos}(-x)=\pi-\operatorname{acos}(x)$.

When comparing the expressions given by Eqs.~\eqref{eq:c_commensurate_EVEN} (for even $q$) and \eqref{eq:c_commensurate_ODD} (for odd $q$), we observe that both share a common underlying structure in which the concentration field is represented as a finite superposition of Green’s function–type kernels of the form $1/\sqrt{\mu-\cos\theta_k}$, reflecting the same geometric distance-based decay from a discrete set of image-like sources. However, a clear distinction arises between the even and odd commensurate cases. For $\alpha=\pi/(2n)$, the solution reduces to a purely algebraic sum with no additional modulation, indicating a higher degree of symmetry and complete cancellation of angular phase contributions. In contrast, for $\alpha=\pi/(2n+1)$, each kernel term is weighted by a nontrivial inverse-cosine factor, which acts as an additional angular mapping that encodes residual geometric asymmetry. Thus, while both cases retain the same fundamental kernel structure, the even case corresponds to a fully symmetric image construction, whereas the odd case exhibits a more complex angular modulation of each contribution.

In particular, for $\alpha = \pi/2$, obtained by setting $n=1$ in Eq.~\eqref{eq:c_commensurate_EVEN} and corresponding to the planar wall limit, the solution reduces to the classical solution 
\begin{equation}
    c  = \frac{1}{2\pi  \sqrt{r^2+\rho^2+2\rho r \sin\theta + z^2}} = \frac{1}{2\pi} \frac{1}{\left| \bm{r}-\bm{r}_0\right|} \, ,
    \label{Eq:c_planar_NN}
\end{equation}
which is proportional to the inverse distance from the singularity located at $(0,\rho,0)$ in Cartesian coordinates.

\begin{figure}
    \centering
    \includegraphics[width=\textwidth]{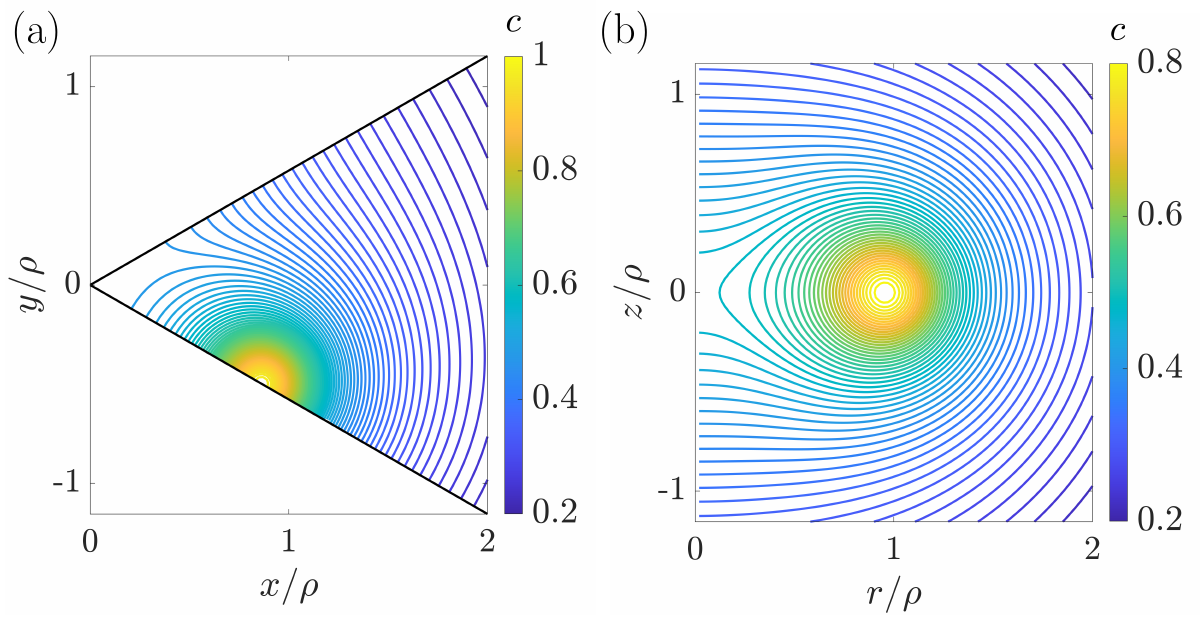}
    \caption{Isoconcentration contours in a wedge geometry for $\alpha=\pi/6$, shown in (a) the radial–azimuthal plane and (b) the radial–axial plane, generated by a point-like active patch on the lower wall, located at $r=\rho$ and $z=0$, with no-flux boundary conditions on the upper wall. In (a) we fix $z/\rho=0.2$, and in (b) $\theta=-\pi/12$.}
    \label{fig:NN_conc}
\end{figure}

\begin{figure}
    \centering
    \includegraphics[width=\textwidth]{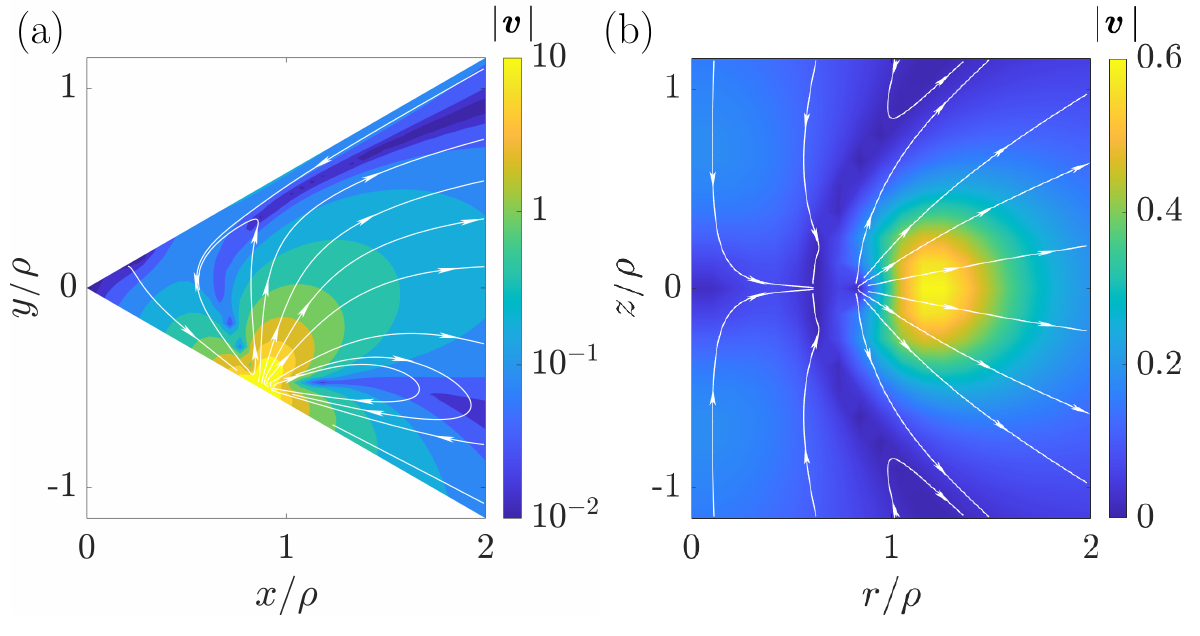}
    \caption{Quiver and contour plots of the magnitude of the phoretic flow velocity induced by a point-like active patch on the lower wall, located at $r=\rho$ and $z=0$, in a wedge geometry for $\alpha=\pi/6$, shown in (a) the radial–azimuthal plane and (b) the radial–axial plane with no-flux boundary conditions on the upper wall. In (a), $z=0$ is fixed, while in (b) $\theta=0$.
    In (a), colours are shown on a logarithmic scale for improved clarity.}
    \label{fig:NN_flow}
\end{figure}

In Fig.~\ref{fig:NN_conc}, we present the isoconcentration contours in a wedge-shaped geometry with semi-opening angle $\alpha=\pi/6$, generated by a point active patch located on the lower wall at the point $(r,\theta,z)=(\rho,-\alpha,0)$, under Neumann–Neumann boundary conditions on both walls. The contours are shown in (a) the radial–azimuthal plane at $z/\rho=0.2$ and (b) the radial–axial plane at $\theta=-\alpha/2$. Lengths are scaled by $\rho$, which denotes the radial position of the patch. Since $\alpha$ corresponds to a commensurate angle with even $q=2n$ ($n=3$), the concentration is evaluated using the exact analytical expression given in Eq.~\eqref{eq:c_commensurate_EVEN}.

From Fig.~\ref{fig:NN_conc}(a), we observe that the wedge geometry induces a pronounced deformation of the isoconcentration contours as a consequence of the no-flux boundary conditions on the walls. This leads to the concentration isolines becoming orthogonal to the wedge boundaries. Far from the wedge surfaces, the concentration field smoothly relaxes to its background value, and the influence of the wedge geometry diminishes. 
Fig.~\ref{fig:NN_conc}(b) shows that the isoconcentration contours on the right-hand side of the plot are only weakly influenced by the wedge. In contrast, close to the wedge apex at $r = 0$, the iso-lines are strongly distorted by the presence of the boundary. Overall, the method presented here demonstrates robustness in computing the concentration field within the wedge geometry, for which an exact analytical solution is available.

Fig.~\ref{fig:NN_flow} shows the phoretic flow induced by the active patch for Neumann–Neumann boundary conditions imposed on the wedge surfaces. 
The point-like active patch is located at $(r,\theta,z)=(\rho,-\alpha,0)$.
The wedge has a semi-opening angle $\alpha=\pi/6$, and results are presented in (a) the radial–azimuthal plane at $z=0$ and (b) the radial–axial plane at $\theta=0$.
As already mentioned, these results are obtained via numerical integration with respect to the radial wavenumber, as an analytical evaluation is not tractable, in contrast to the concentration field case.
In Fig.~\ref{fig:NN_flow}(a), we display the magnitude of the planar velocity field in a logarithmic scale to highlight the large variations in magnitude across the domain, enabling clearer visualization. We observe a flow emerging from the location of the point-like patch, directed predominantly normal to the lower wall, while entering laterally, reminiscent of the flow generated by a dipolar swimmer and characterised by the formation of recirculation zones.
This behavior arises because, along the lower wall at $\theta=-\alpha$, the radial slip velocity satisfies $v_r>0$ for $r<\rho$ and $v_r<0$ for $r>\rho$, with a singularity at $r=\rho$. Along the upper wall at $\theta=\alpha$, the radial slip velocity is predominantly negative and changes sign only near the wedge apex.
In Fig.~\ref{fig:NN_flow}(b), we present the corresponding results in the mid-plane of the wedge, illustrating the induced phoretic flow in the radial–axial cross-section.

\section{Green's function for Neumann–Dirichlet boundary conditions}\label{sec:neumann-dirichlet}

We next examine the case where the phoretic flow is generated by a point activity patch located on the lower wedge at $(r,\theta,z) = (\rho,-\alpha,0)$, while absorbing boundary conditions are imposed on the upper wedge. Compared with Sec.~\ref{sec:neumann}, the key modification is that the upper wall now absorbs solute, leading to a suppressed concentration field at the boundary and a corresponding change in the induced phoretic slip.
This corresponds to a mixed Dirichlet–Neumann boundary condition of the form
\begin{subequations}
    \begin{align}
    -\frac{1}{r}\, \frac{\partial c}{\partial \theta}
    &= \delta(r-\rho) \, \delta(z) \,\,\quad \text{for}\quad \theta=-\alpha \, , \\
    c &= 0 \hspace{6em} \, \text{for}\quad \theta=\alpha \, .
\end{align}
\end{subequations}

\subsection{Solution in FKL space}

We consider a mixed Neumann–Dirichlet boundary-value formulation with $A_-(r,z)=\delta(r-\rho)\,\delta(z)$ and $C_+(r,z)=0$, which yield the FKL transforms $\widetilde{B}_- = \K_{i\nu}(|k|\rho)$ and $\widetilde{C}_+=0$.
Substituting these expressions into Eq.~\eqref{eq:concentration_mixed} yields the concentration field in the form
\begin{equation}
    \widetilde{c} (\nu, \theta, k) = \frac{1}{\nu} \, \sch(2\alpha\nu) \sh \left( (\alpha-\theta)\nu\right) \K_{i\nu}(|k|\rho) \, .
    \label{eq:c_FKL_mixed}
\end{equation}

We express the coefficients in the form given in Eq.~\eqref{eq:coef_form}. For $\upPhi_w$, they are obtained as
\begin{equation}
    \Lambda_w = -\frac{1}{\nu}\, \ch(\alpha\nu)\sch(2\alpha\nu)\sh(\pi\nu) \, , \qquad
    \Lambda_w^\dagger = \frac{1}{\nu}\, \sh(\alpha\nu)\sch(2\alpha\nu)\sh(\pi\nu) \, .
\end{equation}

The remaining coefficients have a more complex structure, given by
\begin{subequations}
    \begin{align}
    \Lambda_x &= \phantom{+} \delta_+ \sin\alpha \ch(\alpha\nu) \left( \Delta_+ - \upGamma\rho\, \tfrac{\partial}{\partial\rho} \right) , \\
    \Lambda_x^\dagger &= -\delta_- \sin\alpha \sh(\alpha\nu) \left( \Delta_- + \upGamma \rho\, \tfrac{\partial}{\partial \rho} \right) , \\
    \Lambda_y &= \phantom{+} \delta_- \cos\alpha \ch(\alpha\nu) \left( \Delta_- + \upGamma \rho\, \tfrac{\partial}{\partial\rho} \right) , \\
    \Lambda_y^\dagger &= -\delta_+ \cos\alpha \sh(\alpha\nu) \left( \Delta_+ -\upGamma \rho\, \tfrac{\partial}{\partial\rho} \right) , 
\end{align}
\end{subequations}
where
\begin{equation}
    \Delta_\pm = 1 \pm \frac{ 2\cos(2\alpha) \ch(2\alpha\nu) }{\ch(4\alpha\nu)+\cos(4\alpha)} \, , \qquad
    \upGamma = \frac{2}{\nu} \frac{\sin(2\alpha) \sh(2\alpha\nu)}{\ch(4\alpha\nu) + \cos(4\alpha)} \, ,
\end{equation}
and $\delta_\pm$ is given by Eq.~\eqref{eq:delta_pm}.

\subsection{Concentration field in real space}

From the expression for the concentration field in FKL space given by Eq.~\eqref{eq:c_FKL_mixed}, the corresponding real-space solution is obtained by applying the inverse FKL transform defined in Eq.~\eqref{eq:c_real_space_def}. This results in a single integral over the radial wavenumber, namely,
\begin{equation}
     c(r,\theta,z) = \frac{1}{2\pi \sqrt{\rho r}}
    \int_0^\infty 
    \sch (2\alpha\nu) \sh \left( (\alpha-\theta)\nu \right) \th(\pi\nu) \,
    \mathrm{P}_{i\nu-\frac{1}{2}}(\mu)
    \, \mathrm{d}\nu \, .
    \label{eq:concentration_mixed}
\end{equation}

To make analytical progress, we use the integral representation of the Legendre function given in Eq.~\eqref{eq:LegendreP_int}, thereby expressing the concentration as a double integral of the form.
\begin{equation}
    c(r,\theta,z) = \frac{1}{\pi^2 \sqrt{2\rho r}}
    \int_{\operatorname{ach}\mu}^\infty 
    \frac{\mathrm{d}t}{\sqrt{\ch t-\mu}}
    \int_0^\infty 
    \sin(\nu t)
    \sch(2\alpha\nu)
    \sh \left( (\alpha-\theta)\nu\right)
     \mathrm{d}\nu \, ,
\end{equation}
where the order of integration has been swapped. 
Absolute convergence permits the interchange of the order of integration.

The integral over the radial wavenumber $\nu$ can be evaluated in closed form, yielding the expression given by
\begin{equation}
    \int_0^\infty 
    \sin(\nu t) \sh \left( (\alpha-\theta)\nu\right)
    \sch(2\alpha\nu) \, \mathrm{d}\nu
    = \frac{\pi}{2\alpha}
    \frac{\cos \left(  \left(1+\frac{\theta}{\alpha} \right) \frac{\pi}{4} \right) 
    \sh \frac{\pi t}{4\alpha} }{\ch \frac{\pi t}{2\alpha} + \sin \frac{\pi\theta}{2\alpha} } \, ;
\end{equation}
see, for instance, Gradshteyn and Ryzhik~\cite[p.~510, Eq.~3.981.6]{gradshteyn2014table}.

As before, the resulting integral with respect to $t$ cannot be evaluated in closed form. To obtain closed-form analytical expressions, we restrict attention to commensurate angles of the form $\alpha=\pi/q$, where $q$ is a positive integer. The cases of even and odd $q$ are then treated separately.

For $q=2n$, we define $\xi_j = (2j-1)\pi/n$, where $j=1,\ldots,m$, with $n=2m$ when $n$ is even and $n=2m+1$ when $n$ is odd. In this case, the concentration field can be expressed as
\begin{equation}
    c(r,\theta,z) = \frac{1}{2\pi^2 \sqrt{\rho r}}
    \left( \cos\tfrac{n\theta}{2} - \sin\tfrac{n\theta}{2} \right)
    \chi_n(\mu,\theta)\, ,
    \label{eq:c_M_even_tmp}
\end{equation}
where \begin{equation}
    \chi_n(\mu,\theta) = \int_{\operatorname{ach}\mu}^\infty 
    \frac{n\sh \frac{nt}{2} }{\ch(nt)+\sin(n\theta)}
    \frac{\mathrm{d}t}{\sqrt{\ch t-\mu}}  \, .
\end{equation}

This integral can be evaluated analytically in closed form.
As shown in the Appendix, the concentration field can be written as a finite sum of terms of the form 
\begin{equation}
    \chi_n(\mu,\theta) = 
    \sum_{k=1}^n X_{kn}(\mu, \theta_k) \, ,
\end{equation}
where the expression for $X_{kn}$ depends on whether $n$ is even or odd; in the case $n=2m$, we obtain
\begin{equation}
    X_{k, 2m}(\mu,\theta_k) =
    \frac{\pi}{2m}
    \sum_{j=1}^m
    (-1)^{j+1} \,
    \frac{\sin\xi_j}{ 
    \cos\xi_j - \cos\theta_k}
    \left( 
    \frac{1}{\sqrt{\mu-\cos\xi_j}}
    -
    \frac{1}{\sqrt{\mu - \cos\theta_k}}
    \right) .
\end{equation}

For $n=2m+1$, the calculations are more involved, but a closed analytical expression is nevertheless possible, yielding
\begin{equation}
X_{k,2m+1} (\mu,\theta_k)
=
\frac{2\sqrt{2}}{2m+1}
\sum_{j=1}^m
(-1)^{j+1} \,
\frac{ \sin\xi_j \cos\frac{\xi_j}{2} }{\cos\theta_k-\cos\xi_j}
\left(
H(\cos\theta_k)
-
H(\cos\xi_j)
\right),
\label{eq:Xk2mP1}
\end{equation}
where
\begin{equation}
     H(x) = \frac{\operatorname{asin}\sqrt{\frac{x+1}{\mu+1}}}{\sqrt{\mu-x}\sqrt{1+x}} \, .
\end{equation}

Finally, for $q=2n+1$, we define $\xi_j = (2j-1)\pi/(2n+1)$, where $j=1,\ldots,n$. In this case, we obtain
\begin{equation}
    c(r,\theta,z) = \frac{1}{2\pi^2 \sqrt{\rho r}}\,
    \left( 
    \cos\left( (2n+1)\tfrac{\theta}{4}\right)
    -
    \sin\left( (2n+1)\tfrac{\theta}{4}\right)
    \right)
    \zeta_n(\mu,\theta) \, ,
    \label{eq:c_M_odd_tmp}
\end{equation}
where
\begin{equation}
    \zeta_n(\mu,\theta) = 
    \bigintsss_{\operatorname{ach}\mu}^\infty 
    \frac{\left( n+\frac12 \right) \sh\left(\left(n+\frac{1}{2} \right)\frac{t}{2}\right)}{\ch\left(\left(n+\frac{1}{2} \right)t\right)+\sin\left(\left(n+\frac{1}{2}\right)\theta\right)}
    \frac{\mathrm{d}t}{\sqrt{\ch t-\mu}} \, .
\end{equation}

As shown in the Appendix, the final solution can be cast in the form
\begin{equation}
    \zeta_n(\mu,\theta) = 
    \sum_{k=1}^{2n+1}
    Z_{kn}(\beta, \theta_k) \, , 
\end{equation}
where, again, $\beta=\sqrt{(\mu+1)/2}$.
In addition, $Z_{kn}$ is given by 
\begin{equation}
    Z_{kn}(\beta,\theta_k) = \frac{2}{2n+1}
    \sum_{j=1}^n (-1)^{j+1}  
    \frac{\sin\xi_j \cos  \frac{\xi_j}{2} }{\cos \frac{\theta_k}{2} - \cos\xi_j}
    \left( Q \left( \cos \tfrac{\theta_k}{2} \right)
    -
    Q \left( \cos\xi_j\right)
    \right) ,
\end{equation}
where 
\begin{equation}
    Q(x) = \frac{1}{\sqrt{2\beta} (1+x)}
    \left( 
    \frac{\beta+1}{\beta-x}\, \upPi\left( -\frac{1+x}{\beta-x} ~\bigg|~ \frac{\beta-1}{2\beta} \right)
    -K \left( \frac{\beta-1}{2\beta} \right)
    \right) .
\end{equation}
Here, $K$ and $\upPi$ denote the complete elliptic integrals of the first and third kind, respectively.

We observe that, for even $q=2n$, the wedge admits a fully closed reflection structure, so the angular eigenmodes come in symmetric pairs and the image construction terminates after a finite number of reflections. As a result, the concentration field reduces to a finite discrete sum with purely algebraic dependence on $\mu$ and $\theta$, involving only trigonometric functions and square-root singularities. Physically, transport is fully captured by a finite set of mirror contributions, with no residual spectral coupling.
For odd $q=2n+1$, the reflection symmetry does not close and the angular modes do not fully pair under the wedge reflections. The image construction therefore does not reduce entirely to a finite algebraic structure, leaving a residual nonlocal contribution in the spectral representation. This manifests as elliptic integrals in the closed form, reflecting the incomplete cancellation of reflected contributions and a more complex angular coupling in the concentration field.

\begin{figure}
    \centering
    \includegraphics[width=\textwidth]{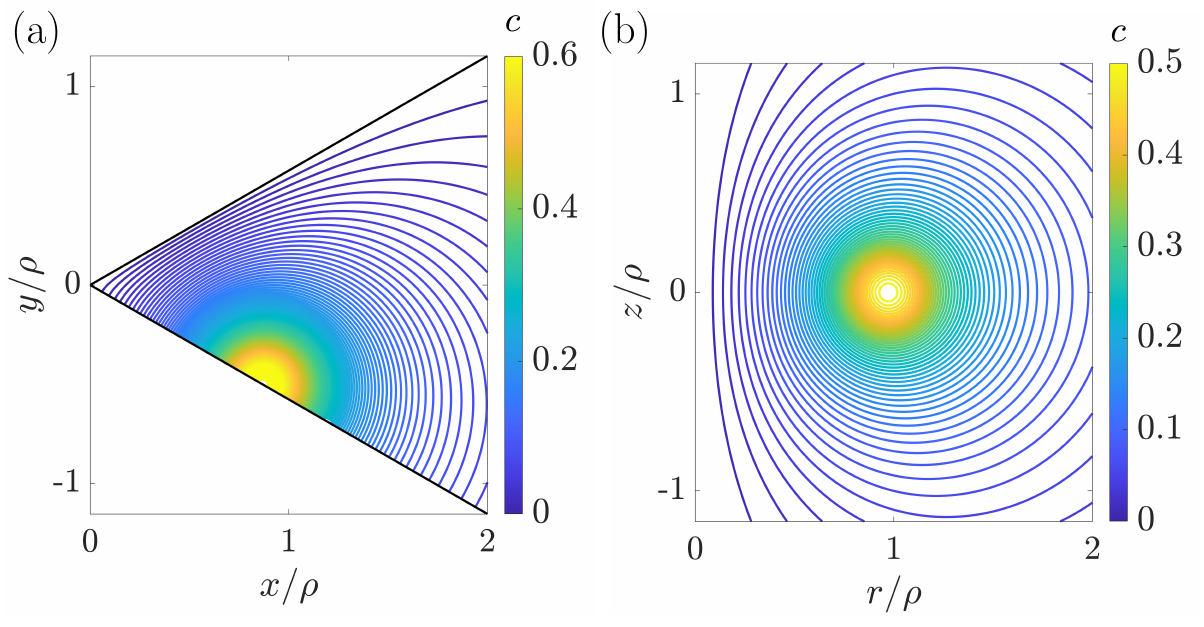}
    \caption{Isoconcentration contours in a wedge geometry for $\alpha = \pi/6$, shown in (a) the radial–azimuthal plane and (b) the radial–axial plane, generated by a point active patch on the lower wall, located at $r=\rho$ and $z=0$, with absorbing boundary conditions imposed on the upper wall. In (a), we fix $z/\rho = 0.2$, while in (b), $\theta = -\pi/12$.
}
    \label{fig:M_conc}
\end{figure}

In Fig.~\ref{fig:M_conc}, we show contours of equal concentration in a wedge-shaped geometry with semi-opening angle $\alpha=\pi/6$, generated by a point active patch located on the lower wall at $(r,\theta,z)=(\rho, -\alpha,0)$, under mixed boundary conditions with a Neumann condition on the lower wall and an absorbing condition on the upper wall. The contours are displayed in (a) the radial–azimuthal plane and (b) the radial–axial plane, using the same parameters as in Fig.~\ref{fig:NN_conc}. Since $\alpha$ corresponds to a commensurate angle with even $q=2n$, where $n=2m+1$ is odd and $m=1$, the concentration is evaluated using the exact analytical expression in Eq.~\eqref{eq:c_M_even_tmp}, with $X_{kn}$ given by Eq.~\eqref{eq:Xk2mP1}.

From the absorbing boundary condition on the upper wall ($c=0$), the isoconcentration contours in Fig.~\ref{fig:M_conc}(a) exhibit a strong asymmetry, with lines terminating at the absorbing surface and being drawn toward the sink-like boundary, in contrast to the symmetric deformation observed under no-flux conditions shown in Fig.~\ref{fig:NN_conc}(a). The concentration field is significantly depleted near the upper wall.
Fig.~\ref{fig:M_conc}(b) shows that the wedge geometry exerts its strongest influence near the apex, where the isolines undergo pronounced distortion.

\begin{figure}
    \centering
    \includegraphics[width=\textwidth]{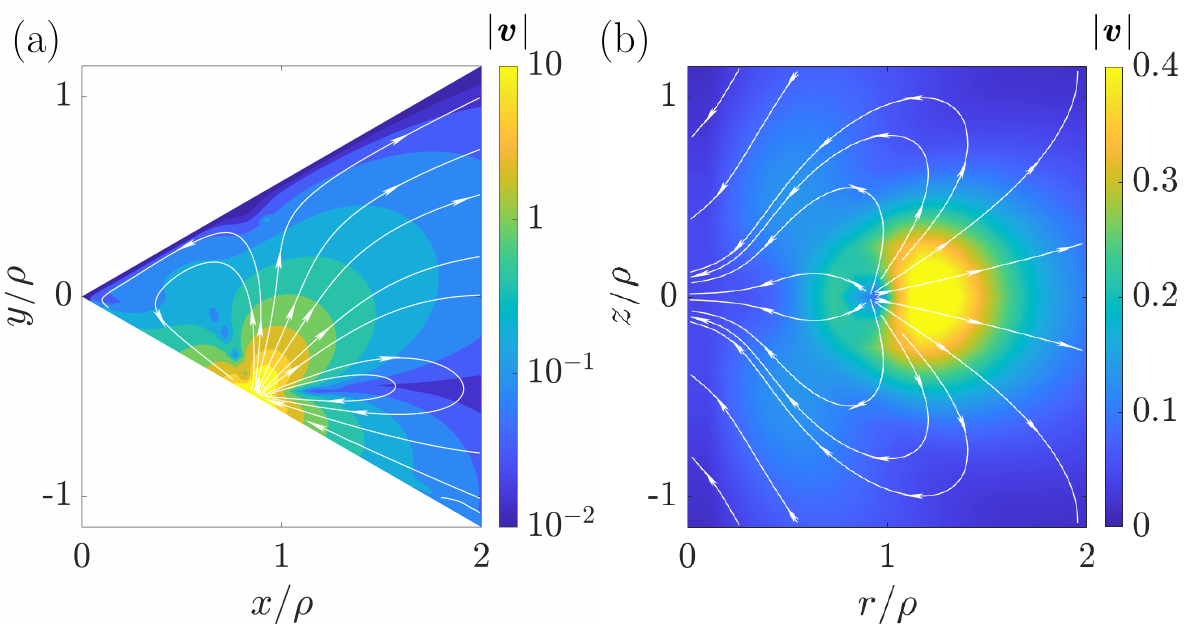}
    \caption{Quiver and contour plots illustrating the magnitude of the phoretic flow velocity induced by a point-like active patch on the lower wall, located at $r=\rho$ and $z=0$, in a wedge geometry for $\alpha=\pi/6$, shown in (a) the radial–azimuthal plane and (b) the radial–axial plane with absorbing boundary conditions imposed on the upper wall. In (a), $z=0$ is fixed, while in (b), $\theta=0$.
    The colours in (a) are displayed on a logarithmic scale for clarity.
}
\label{fig:M_flow}
\end{figure}

In Fig.~\ref{fig:M_flow}, we present the phoretic flow generated by the active patch under mixed Neumann–Dirichlet boundary conditions imposed on the wedge surfaces. 
The point-like active patch is located at $(r,\theta,z)=(\rho,-\alpha,0)$.
The wedge has a semi-opening angle of $\alpha=\pi/6$, and the flow is shown in (a) the radial–azimuthal plane at $z=0$ and (b) the radial–axial plane at $\theta=0$. These results are obtained by numerically integrating over the radial wavenumber.
Figure~\ref{fig:M_flow}(a) displays the magnitude of the planar velocity field on a logarithmic scale. The flow emerges from the point-like active patch and is directed predominantly normal to the lower wall while entering laterally, resembling the flow generated by a force-dipole swimmer and exhibiting characteristic recirculation zones. Along the lower wall, the fluid is drawn tangentially towards the patch and expelled in the wall-normal direction. In contrast, the radial slip velocity vanishes along the upper wall, producing a streamline pattern that differs markedly from that obtained under Neumann–Neumann boundary conditions in Fig.~\ref{fig:NN_flow}(a).
The corresponding flow in the radial–axial cross-section is shown in Fig.~\ref{fig:M_flow}(b), where the velocity field is displayed in the mid-plane of the wedge, illustrating the three-dimensional structure of the induced phoretic flow.

\section{Conclusions}\label{sec:conclusions}

In this work, we have developed a theoretical framework for phoretically driven flows in a three-dimensional wedge geometry. The problem is motivated by the use of chemically active surfaces to generate solute gradients and, through diffusioosmotic slip, drive bulk Stokes flows without externally imposed pressure gradients or moving boundaries~\cite{anderson1989,michelin2014}. In the limit of small Péclet and Reynolds numbers, the transport problem becomes unidirectionally coupled: the solute concentration is first determined from a harmonic boundary-value problem, and its tangential gradients subsequently prescribe the slip velocity driving the bulk hydrodynamic flow. This setting extends the recently studied two-dimensional diffusioosmotic corner flows~\cite{Nowak_Lisicki_2026}, but the fully three-dimensional geometry requires a different mathematical formulation because the scalar stream-function representation is no longer available.

To address this challenge, we combine the Fourier--Kontorovich--Lebedev (FKL) transform with the Papkovich--Neuber representation of Stokes flow~\cite{kontorovich1938one,lebedev1949,Papkovich1932,Neuber1934,tran1982general,daddi2026spectral}. The FKL transform is particularly well suited to wedge geometries with translational invariance along the edge, as it separates the axial and radial dependences and reduces the Laplace problem to an ordinary differential equation in the polar coordinate. The Papkovich--Neuber representation then expresses the hydrodynamic problem in terms of harmonic potentials, preserving the spectral structure inherited from the concentration field. This approach enables the chemically generated scalar field and the induced Stokes flow to be treated within a unified analytical framework.

Using this formulation, we derive Green's functions for two physically relevant boundary-value problems. In the Neumann--Neumann case, a point-like active patch on one wedge face is coupled to a no-flux condition on the opposite wall, whereas in the mixed Neumann--Dirichlet case the opposite wall acts as an absorbing boundary. For arbitrary wedge opening angles, the concentration fields are expressed as single integrals over the radial spectral variable. When the semi-opening angle is commensurate, $\alpha=\pi/q$, these representations can be further evaluated to obtain closed analytical expressions. For the Neumann--Neumann problem, the resulting solutions reduce to finite image-like sums involving inverse-square-root kernels, with distinct forms for even and odd values of $q$. In the mixed problem, the corresponding expressions exhibit a more intricate angular dependence and, in certain cases, involve complete elliptic integrals. These analytical solutions provide valuable benchmarks for numerical studies of diffusion in three-dimensional wedge domains.

We obtain the induced viscous flow from the corresponding Papkovich--Neuber potentials. While the concentration field admits analytical evaluation for several commensurate-angle configurations, the velocity field generally requires numerical integration over the spectral variable. This distinction arises from the increased complexity of the hydrodynamic boundary conditions: the slip velocity depends on tangential concentration gradients, while the Papkovich--Neuber coefficients involve non-local spectral couplings. The numerical examples demonstrate that a localised active patch generates a genuinely three-dimensional flow, with fluid expelled in the wall-normal direction and drawn towards the active region along the surfaces. The resulting recirculating flow structure resembles a dipolar Stokes field and is strongly influenced by the boundary condition imposed on the opposite wedge face. In particular, an absorbing wall produces a qualitatively different streamline topology compared with a no-flux boundary, as the concentration field and the resulting slip velocity are modified throughout the domain.

The Green's function nature of the solutions makes the framework directly applicable to more general activity distributions. Finite catalytic patches, multiple active regions, and patterned surfaces can all be constructed through superposition of the point-source solutions derived here. The theory therefore provides not only an exact solution to a canonical three-dimensional Stokes-flow problem but also a design and benchmarking tool for phoretic pumping and mixing in cornered microfluidic geometries, dead-end pores, and chemically patterned confined systems~\cite{michelin2015geometric,michelin2019universal,Yu2020,Visan2024}.

Several extensions naturally follow from the present formulation. First, the point-like activity considered here can be regularised into finite patches, providing a more realistic representation of chemically active surfaces and enabling direct comparison with experimental systems. Such finite patches would allow one to investigate how the size, shape, and spatial distribution of the active regions control the resulting solute gradients, phoretic slip, and induced flow. Recent advances in fabrication and surface-functionalisation techniques allow catalytic, enzymatic, and electrostatic activity to be patterned with increasing spatial precision~\cite{Archer2015,Kreienbrink2025,Sengupta2014,Stroock2000,stroock2003}, opening new possibilities for exploiting phoretic mechanisms in microfluidic applications. Second, spatial variations in phoretic mobility could be incorporated alongside heterogeneous activity, allowing surfaces patterned both chemically and physicochemically to be investigated. Third, finite Péclet-number effects would introduce feedback from the flow onto the solute distribution, resulting in a nonlinear advection--diffusion--flow problem~\cite{michelin2014}. Incorporating finite wedge lengths, time-dependent activity, reactive surface kinetics, or electrolyte-specific interactions would bring the model closer to experimentally realistic phoretic systems. Such extensions could provide insight into applications including nanoscale diffusioosmotic pumping~\cite{Chanda2022}, guided transport into microcavities~\cite{VrhovecHartman2018}, intensity-controlled photocatalysis~\cite{Timmerhuis}, enzyme-driven activity~\cite{Popescu2025}, and light-activated diffusioosmotic flows~\cite{Muraveva2024}. The analytical structure developed here provides a foundation for pursuing these directions while retaining the geometric advantages of the FKL representation.
Finally, an extension to Dirichlet–Dirichlet boundary conditions, which would also give access to the corresponding hydrodynamic problem with prescribed concentration, is another interesting direction for future work.

\enlargethispage{20pt}

\ack{M.L.\ thanks A.D.M.I.\ and the Department of Mathematics and Statistics at The Open University, Milton Keynes, for their hospitality during a visit in which this work was initiated. The work was supported by the National Science Centre of Poland (Sonata Bis grant no.\ 2023/50/E/ST3/00465 awarded to M.L.).}

\appendix

\section{Derivation of the integral identities}\label{sec:appA}

In this Appendix, we derive the integral identity that leads to the series representation of the concentration fields. These integrals are highly non-trivial and, in general, cannot be evaluated directly using standard computer algebra systems such as Mathematica or Maple. However, by applying a sequence of suitable transformations, an analytical evaluation becomes possible.

\subsection{Integrals related to the solution for Neumann-Neumann boundary conditions}

\subsubsection{Evaluation of the integral in the case of even~$q$}

We consider the integral appearing in Eq.~\eqref{eq:c_NN_even_tmp}, corresponding to the case $q=2n$. We define the improper integral
\begin{equation}
    \phi_n(\mu, \theta)
    =
    \int_{\operatorname{ach}\mu}^\infty 
    \frac{n \sh(nt)}{\ch(nt)+\sin(n\theta)}
    \frac{\mathrm{d}t}{\sqrt{\ch t-\mu}} \, ,
    \label{eq:appendix_int1_def}
\end{equation}
where $\mu \ge 1$.
We have
\begin{equation}
    \ch(nt)+\sin(n\theta) =
    2^{n-1} \prod_{k=1}^n \left( \ch t-\cos\theta_k \right) ,
    \label{eq:sum_to_prod_Tn}
\end{equation}
with $\theta_k = \theta - \left( 4k -1\right) \pi/(2n)$.
This identity is obtained by exploiting the fact that the left-hand side can be viewed as a function of $\ch t$ via the relation $\ch(nt)=T_n(\ch t)$, where $T_n$ denotes the Chebyshev polynomial of the first kind, and then applying the standard result that a polynomial admits a factorization into linear factors corresponding to its roots. In this formulation, the roots are located at the Chebyshev nodes $\cos\theta_k$, which leads directly to the product representation on the right-hand side.

By taking the logarithm of both sides of Eq.~\eqref{eq:sum_to_prod_Tn} and differentiating with respect to $t$, we obtain a series representation by converting the product structure into a sum over its linear factors.
Specifically, 
\begin{equation}
    \frac{n\sh(nt)}{\ch(nt)+\sin(n\theta)}
    = 
    \sum_{k=1}^n 
    \frac{\sh t}{\ch t - \cos \theta_k} \, .
    \label{eq:ser_rep_appendix_even}
\end{equation}

The integral in Eq.~\eqref{eq:appendix_int1_def} can now be evaluated by introducing the change of variables $v = \sqrt{\ch t - \mu}$. Hence,
\begin{equation}
    \phi_n(\mu,\theta) = \sum_{k=1}^n \int_0^\infty
    \frac{2\, \mathrm{d}v}{v^2+\mu-\cos\theta_k} \, ,
\end{equation}
which yields the desired result
\begin{equation}
    \phi_n(\mu, \theta) = 
    \sum_{k=1}^{n} \frac{\pi}{\sqrt{\mu - \cos \theta_k}}\, .
\end{equation}

\subsubsection{Evaluation of the integral in the case of odd~$q$}

We next derive the integral appearing in Eq.~\eqref{eq:c_NN_odd_tmp}, which applies in the case $q = 2n + 1$. We define
\begin{equation}
\psi_n(\mu,\theta) =
    \bigintsss_{\operatorname{ach}\mu}^\infty 
    \frac{\left( n+\frac12 \right) \sh\left(\left(n+\frac{1}{2} \right)t\right)}{\ch\left(\left(n+\frac{1}{2} \right)t\right)+\sin\left(\left(n+\frac{1}{2}\right)\theta\right)}
    \frac{\mathrm{d}t}{\sqrt{\ch t-\mu}} \, .
\end{equation}

We use the series representation
\begin{equation}
    \frac{\left( n+\frac12 \right) \sh\left(\left(n+\frac{1}{2} \right)t\right)}{\ch\left(\left(n+\frac{1}{2} \right)t\right)+\sin\left(\left(n+\frac{1}{2}\right)\theta\right)}
    = \frac{1}{2}
    \sum_{k=1}^{2n+1} \frac{ \sh \frac{t}{2} }{ \ch \frac{t}{2} -\cos \frac{\theta_k}{2} } \, , 
    \label{eq:ser_rep_appendix_odd}
\end{equation}
where $\theta_k = \theta - (4k-1)\, \pi/(2n+1)$.
This follows directly from Eq.~\eqref{eq:ser_rep_appendix_even} upon appropriate changes of variables.
Alternatively, this is obtained by taking the logarithmic derivative of the corresponding product representation with respect to $t$. In particular, the product form is first expressed in terms of linear factors in $\ch(t/2)$, and then differentiation of the logarithm converts the product into a sum over simple poles associated with the zeros $\cos(\theta_k/2)$. This yields a partial fraction expansion, which directly leads to the stated series representation.

We then introduce the change of variable $u=\ch (t/2)$.
Therefore,
\begin{equation}
    \psi_n(\mu,\theta) = 
    \frac{1}{\sqrt{2}}
    \sum_{k=1}^{2n+1} 
    \bigintssss_\beta^\infty \frac{\mathrm{d}u}{\left( u-\cos \frac{\theta_k}{2} \right) \sqrt{u^2-\beta^2}} \, ,
\end{equation}
where $\beta = \sqrt{(\mu+1)/2}$, with $\beta\ge 1$.
Finally, by introducing the change of variable
\begin{equation}
    v=\sqrt{\frac{u-\beta}{u+\beta}} \, ,
\end{equation}
the integral $\psi_n$ can be rewritten in the form of an equivalent integral given by
\begin{equation}
    \int_0^1 \frac{ \mathrm{d}v }{a_+ + a_-v^2} 
    = 
    \frac{1}{\sqrt{a_+a_-}}\, \operatorname{atan} \sqrt{ \frac{a_+}{a_-} } , 
\end{equation}
where $a_\pm = \left( \beta\pm\cos(\theta_k/2) \right)/\sqrt{2}$.
This yields, after simplification, 
\begin{equation}
 \psi_n(\mu,\theta) =
    \sum_{k=1}^{2n+1}
    \frac{1}{\sqrt{\mu-\cos\theta_k}}
    \operatorname{acos} \left( -\tfrac{1}{\beta} \, \cos \tfrac{\theta_k}{2} \right) .
\end{equation}


\subsection{Integrals related to the solution for Neumann-Dirichlet boundary conditions}

\subsubsection{Evaluation of the integral in the case of even~$q$}

We consider the improper integral in Eq.~\eqref{eq:c_M_even_tmp}, corresponding to the case $q=2n$. We introduce the integral
\begin{equation}
    \chi_n(\mu,\theta) = \int_{\operatorname{ach}\mu}^\infty 
    \frac{n\sh\left( \frac{nt}{2} \right)}{\ch(nt)+\sin(n\theta)}
    \frac{\mathrm{d}t}{\sqrt{\ch t-\mu}}  \, .
\end{equation}

Using the trigonometric identity $\sin t = 2\sin(t/2)\cos(t/2)$, it follows from Eq.~\eqref{eq:ser_rep_appendix_even} that the series representation can be rewritten as 
\begin{equation}
    \frac{n\sh \frac{nt}{2} }{\ch(nt)+\sin(n\theta)}
    = 
    \frac{1}{2}\, 
    \sch \frac{nt}{2}
    \sum_{k=1}^n 
    \frac{\sh t}{\ch t - \cos \theta_k} \, , 
    \label{eq:ser_rep_appendix_even_mixed}
\end{equation}

Using the change of variable $v=\sqrt{\ch t-\mu}$, we obtain
\begin{equation}
    \chi_n(\mu, \theta) = \sum_{k=1}^n X_{kn}(\mu, \theta_k) \, , 
\end{equation}
where
\begin{equation}
    X_{kn}(\mu,\theta_k) =
    \int_0^\infty  
    \frac{W_n(v^2+\mu) \, \mathrm{d}v}{v^2+\mu-\cos\theta_k} \, .
    \label{eq:X_kn}
\end{equation}
Here,
\begin{equation}
    W_n(x) = \sch \left( \frac{n}{2}\, \operatorname{ach} x \right)
    = 
    \begin{cases}
         \qquad\,\, 1 /T_{m}(x) & \text{for} \quad n=2m  \, ,   \\[3pt]
        \sqrt{2}/\sqrt{1+T_{2m+1}(x)} & \text{for} \quad n=2m+1 \, .
    \end{cases}
    \label{eq:W_n}
\end{equation}

Thus, depending on the parity of $n$, we treat each case separately. For $n=2m$, the denominator in the integral in Eq.~\eqref{eq:X_kn} is a polynomial of degree $2m+2$, with roots given by $v=\pm i\sqrt{\mu-\cos\theta_k}$ and $v=\pm i\sqrt{\mu-\cos\xi_j}$, where $\xi_j=(2j-1)\pi/(2m)$, so that $\cos\xi_j$ denote the Chebyshev nodes. By applying the residue theorem and closing the contour in the upper half complex plane, we readily obtain
\begin{equation}
    X_{k, 2m}(\mu,\theta_k) =
    \frac{\pi}{2m}
    \sum_{j=1}^m
    (-1)^{j+1} \,
    \frac{\sin\xi_j}{ 
    \cos\xi_j - \cos\theta_k}
    \left( 
    \frac{1}{\sqrt{\mu-\cos\xi_j}}
    -
    \frac{1}{\sqrt{\mu - \cos\theta_k}}
    \right) .
\end{equation}

For $n=2m+1$, the situation is more delicate, since the denominator is no longer a polynomial and involves a square-root structure. In this case, we introduce $\xi_j=(2j-1)\pi/\left(2m+1\right)$, so that $\cos\xi_j$ are the roots of $1+T_{2m+1}$. Accordingly, 
\begin{equation}
    X_{k, 2m+1}(\mu, \theta_k) = 
    \frac{\sqrt{2}}{2^{m}} 
    \int_0^\infty
    \frac{\mathrm{d} v}{\left( v^2+\mu-\cos\theta_k\right) \prod_{j=1}^{2m+1} \sqrt{v^2+\mu-\cos\xi_j}} \, .
\end{equation}
Using the symmetry $\cos\xi_{j+1}=\cos\xi_{2m+1-j}$, the product in the denominator simplifies significantly, leaving only a single square-root contribution, namely
\begin{equation}
    X_{k, 2m+1}(\mu, \theta_k) = 
    \frac{\sqrt{2}}{2^{m}} 
    \int_0^\infty
    \frac{\mathrm{d}v}{\left( v^2+\mu-\cos\theta_k\right) 
    \sqrt{v^2+\mu+1}
    \prod_{j=1}^{m} \left( v^2+\mu-\cos\xi_j\right)} \, .
\end{equation}

Using partial fraction decomposition, this expression can be rewritten as a finite sum of the form
\begin{equation}
    X_{k, 2m+1}(\mu, \theta_k) = 
    \frac{2\sqrt{2} }{2m+1}
    \int_0^\infty
    \frac{\mathrm{d} v}{\left( v^2+\mu-\cos\theta_k\right) 
    \sqrt{v^2+\mu+1}} 
    \sum_{j=1}^{m} (-1)^{j+1}  \frac{\sin\xi_j \cos \frac{\xi_j}{2} }{v^2+\mu-\cos\xi_j} .
\end{equation}

Although the integral may appear daunting at first glance, it admits an antiderivative that can be obtained systematically using computer algebra systems such as Mathematica.
We finally obtain
\begin{equation}
X_{k,2m+1} (\mu,\theta_k)
=
\frac{2\sqrt{2}}{2m+1}
\sum_{j=1}^m
(-1)^{j+1} \,
\frac{ \sin\xi_j \cos\frac{\xi_j}{2} }{\cos\theta_k-\cos\xi_j}
\left(
H(\cos\theta_k)
-
H(\cos\xi_j)
\right),
\end{equation}
where
\begin{equation}
    H(x) = \frac{\operatorname{asin}\sqrt{\frac{x+1}{\mu+1}}}{\sqrt{\mu-x}\sqrt{1+x}} \, .
\end{equation}
Accordingly, for odd $n$, the solution can be written as a double finite sum.

\subsubsection{Evaluation of the integral in the case of odd~$q$}

We next consider the integral in Eq.~\eqref{eq:c_M_odd_tmp}, corresponding to the case $q = 2n + 1$. We introduce
\begin{equation}
    \zeta_n(\mu,\theta) = 
    \bigintsss_{\operatorname{ach}\mu}^\infty 
    \frac{\left( n+\frac12 \right) \sh\left(\left(n+\frac{1}{2} \right)\frac{t}{2}\right)}{\ch\left(\left(n+\frac{1}{2} \right)t\right)+\sin\left(\left(n+\frac{1}{2}\right)\theta\right)}
    \frac{\mathrm{d}t}{\sqrt{\ch t-\mu}} \, .
\end{equation}

Using the sine double-angle formula, it follows from Eq.~\eqref{eq:ser_rep_appendix_odd} that
\begin{equation}
    \frac{\left( n+\frac12 \right) \sh\left(\left(n+\frac{1}{2} \right)\frac{t}{2}\right)}{\ch\left(\left(n+\frac{1}{2} \right)t\right)+\sin\left(\left(n+\frac{1}{2}\right)\theta\right)}
    = 
    \frac{1}{4} \,
    \sch\left(\left(n+\tfrac{1}{2} \right)\tfrac{t}{2}\right)
    \sum_{k=1}^{2n+1} 
    \frac{ \sh \frac{t}{2} }{ \ch \frac{t}{2} -\cos \frac{\theta_k}{2} } \, , 
\end{equation}

Using the change of variable $v=\sqrt{\ch(t/2)-\beta}$, we arrive at
\begin{equation}
    \zeta_n(\mu,\theta) =  \sum_{k=1}^{2n+1}
    Z_{kn}(\beta, \theta_k) \, , 
\end{equation}
where, again, $\beta=\sqrt{(\mu+1)/2}$, and
\begin{equation}
    Z_{kn}(\beta, \theta) =  
    \frac{1}{\sqrt{2}}
    \int_0^\infty 
    \frac{W_{2n+1}\left( v^2+\beta\right)\, \mathrm{d}v}{\left( v^2+N-\cos \frac{\theta_k}{2} \right) \sqrt{v^2+2\beta} } \, , 
\end{equation}
with $W_n$ defined in Eq.~\eqref{eq:W_n}, with the odd-index case applying here.
Specifically,
\begin{equation}
    Z_{kn}(\beta, \theta_k) =  
    \bigintssss_0^\infty 
    \frac{\mathrm{d}v}{\left( v^2+\beta-\cos \frac{\theta_k}{2} \right) \sqrt{v^2+2\beta} 
    \sqrt{1+T_{2n+1}\left( v^2+\beta\right)}
    } \, . 
\end{equation}

As before, we employ a linear factorization of the polynomial under the square root, including Chebyshev polynomials, and by exploiting the symmetry of the product, we obtain
\begin{equation}
Z_{kn}(\beta, \theta_k) =
\frac{1}{2^n}
\bigintssss_0^\infty
\frac{\mathrm{d}v}{\left(v^2 + \beta - \cos\left( \frac{\theta_k}{2}\right)\right)
\sqrt{v^2 + 2\beta}
\sqrt{v^2 + \beta + 1}
\prod_{j=1}^{n}\bigl(v^2 + \beta - \cos(\xi_j)\bigr)} \, , 
\end{equation}
where $\xi_j = (2j-1)\pi/(2n+1)$.

Using partial fraction decomposition, which allows the rational function to be rewritten as a sum of simpler terms associated with its poles, we obtain
\begin{equation}
    Z_{kn}(\beta,\theta_k) =\frac{2}{2n+1}
    \bigintssss_0^\infty 
    \frac{\mathrm{d}v}{\left(v^2 + \beta - \cos \frac{\theta_k}{2}\right)
\sqrt{v^2 + 2\beta}
\sqrt{v^2 + \beta + 1}}
 \sum_{j=1}^{n} (-1)^{j+1}  \frac{\sin\xi_j \cos \frac{\xi_j}{2} }{v^2+\beta-\cos\xi_j} \, .
\end{equation}

This integral can be evaluated by first performing a partial fraction decomposition with respect to the two quadratic factors in the denominator. Then
\begin{equation}
    Z_{kn}(\beta,\theta_k) = \frac{2}{2n+1}
    \sum_{j=1}^n (-1)^{j+1}  
    \frac{\sin\xi_j \cos \frac{\xi_j}{2} }{\cos \frac{\theta_k}{2} - \cos\xi_j}
    \left( Q \left( \cos \tfrac{\theta_k}{2} \right)
    -
    Q \left( \cos\xi_j\right)
    \right) ,
\end{equation}
where 
\begin{equation}
    Q(x) = \int_0^\infty
    \frac{\mathrm{d}v}{\left( v^2+\beta-x\right)
    \sqrt{v^2+2\beta} \sqrt{v^2+\beta+1} } \, .
\end{equation}
The latter integral can be expressed in terms of elliptic integrals as
\begin{equation}
    Q(x) = \frac{1}{\sqrt{2\beta} (1+x)}
    \left( 
    \frac{\beta+1}{\beta-x}\, \upPi\left( -\frac{1+x}{\beta-x} ~\bigg|~ \frac{\beta-1}{2M} \right)
    -K \left( \frac{\beta-1}{2\beta} \right)
    \right) .
\end{equation}
Here, $K$ and $\upPi$ denote the complete elliptic integrals of the first and third kind, respectively.
We note that here we use the convention implemented in Wolfram Mathematica, where the elliptic parameter is defined as the square of the modulus.

For each integral derived above in closed analytical form, its accuracy has been validated by direct comparison with numerical evaluations for both parities of $q$. While numerical integration can be computationally expensive due to the complicated nature of the integrands and the presence of an infinite integration domain, the exact series representations obtained in this work offer a powerful and efficient approach, enabling rapid and accurate evaluation of the integrals.


\vskip2pc

\bibliographystyle{RS}


\end{document}